\documentclass[longauth]{aa}  

\usepackage{graphicx}
\usepackage{eucal}
\usepackage[flushleft]{threeparttable}
\usepackage[utf8]{inputenc}

\usepackage{etoolbox}
\newtoggle{authlist}
\usepackage{multirow}
\usepackage{graphicx}	% Including figure files
\usepackage{amsmath}	% Advanced maths commands
\usepackage{scalerel}
\usepackage{tikz}
\usetikzlibrary{svg.path}
\definecolor{orcidlogocol}{HTML}{A6CE39}
\tikzset{
  orcidlogo/.pic={
    \fill[orcidlogocol] svg{M256,128c0,70.7-57.3,128-128,128C57.3,256,0,198.7,0,128C0,57.3,57.3,0,128,0C198.7,0,256,57.3,256,128z};
    \fill[white] svg{M86.3,186.2H70.9V79.1h15.4v48.4V186.2z}
                 svg{M108.9,79.1h41.6c39.6,0,57,28.3,57,53.6c0,27.5-21.5,53.6-56.8,53.6h-41.8V79.1z M124.3,172.4h24.5c34.9,0,42.9-26.5,42.9-39.7c0-21.5-13.7-39.7-43.7-39.7h-23.7V172.4z}
                 svg{M88.7,56.8c0,5.5-4.5,10.1-10.1,10.1c-5.6,0-10.1-4.6-10.1-10.1c0-5.6,4.5-10.1,10.1-10.1C84.2,46.7,88.7,51.3,88.7,56.8z};
  }
}
\newcommand\orcidicon[1]{\iftoggle{authlist}{%
\textsuperscript{\href{https://orcid.org/#1}{\mbox{\scalerel*{
\begin{tikzpicture}[xscale=0.1,yscale=-0.1,transform shape]
\pic{orcidlogo};
\end{tikzpicture}
}{|}}}}
}{
}}

\usepackage{hyperref}   %Needs to be at the end for some reason

\usepackage{txfonts}
\newcommand{\tess}{{\it TESS}}

\newcommand{\gaia}{{\it Gaia}}

\newcommand{\ktwo}{{\it K2}}
\newcommand{\cheops}{{\it CHEOPS}}

\newcommand{\kms}{km\,s$^{-1}$}
\newcommand{\ms}{m\,s$^{-1}$}

\newcommand{\teff}{$T_{\rm eff}$}

\newcommand{\logg}{$\log g$}

\newcommand{\TICstar}{TIC-27491137}
\newcommand{\monotools}{\texttt{MonoTools}}
\newcommand{\celerite}{\textit{celerite}}
\newcommand{\refcomb}{}
\newcommand{\refcom}{}

\newcommand{\Ttransittimesonezero}{$ 1940.4798 \pm 0.0011 $}

\newcommand{\Ttransittimesoneone}{$ 2274.08398 \pm 0.0008 $}

\newcommand{\Ttransittimesonetwo}{$ 1938.2915 \pm 0.0014 $}

\newcommand{\Trorzero}{$ 0.02998^{+0.00035}_{-0.00035} $}

\newcommand{\Trorone}{$ 0.04164 \pm 0.0004 $}

\newcommand{\Trortwo}{$ 0.03848 \pm 0.00069 $}

\newcommand{\TRs}{$ 0.7699 \pm 0.0059 $}

\newcommand{\Trplzero}{$ 2.518 \pm 0.036 $}

\newcommand{\Trplone}{$ 3.497 \pm 0.043 $}

\newcommand{\Trpltwo}{$ 3.232 \pm 0.063 $}

\newcommand{\Tsmazero}{$ 0.0682 \pm 0.0013 $}

\newcommand{\Tsmaone}{$ 0.1093 \pm 0.0021 $}

\newcommand{\Tsmatwo}{$ 0.1539 \pm 0.0029 $}

\newcommand{\TSinzero}{$ 114400^{+6900}_{-6500} $}

\newcommand{\TSinone}{$ 44500 \pm 2600 $}

\newcommand{\TSintwo}{$ 22400 \pm 1300 $}

\newcommand{\TTsurfpzero}{$ 797.0 \pm 12.0 $}

\newcommand{\TTsurfpone}{$ 629.5 \pm 9.2 $}

\newcommand{\TTsurfptwo}{$ 530.4 \pm 7.8 $}

\newcommand{\Ttdurzero}{$ 3.251 \pm 0.03 $}

\newcommand{\Ttdurone}{$ 4.186 \pm 0.029 $}

\newcommand{\Ttdurtwo}{$ 3.046 \pm 0.047 $}

\newcommand{\Teccb}{ $ 0.023 \pm 0.02 $ }

\newcommand{\Teccc}{ $ 0.047^{+0.028}_{-0.024} $ }

\newcommand{\Teccd}{ $ 0.075 \pm 0.052 $ }

\newcommand{\TttvMb}{ $ 5.9 \pm 2.8 $ }

\newcommand{\TttvMc}{ $ 6.4 \pm 2.9 $ }
\newcommand{\TttvMcshort}{ $ 6.0 $ }

\newcommand{\TttvMd}{ $ 6.7^{+4.5}_{-2.9} $ }

\newcommand{\Tttvperb}{ $ 10.35509^{+0.0002}_{-0.00014} $ }
\newcommand{\Tttvperbshort}{ $ 10.3551 $ }

\newcommand{\Tttvperc}{ $ 21.01538^{+0.00084}_{-0.00074} $ }
\newcommand{\Tttvpercshort}{ $ 21.0154 $ }

\newcommand{\Tttvperd}{ $ 35.12537 \pm 0.00067 $ }

\newcommand{\TttvKb}{ $ 1.88 \pm 0.87 $ }

\newcommand{\TttvKc}{ $ 1.62 \pm 0.71 $ }

\newcommand{\TttvKd}{ $ 1.45^{+0.96}_{-0.61} $ }

\newcommand{\TttvTSMb}{ $ 150^{+130}_{-50} $ }

\newcommand{\TttvTSMc}{ $ 280^{+220}_{-90} $ }

\newcommand{\TttvTSMd}{ $ 180 \pm 100 $ }

\newcommand{\Tttvbcsuperper}{ $ 713.1 \pm 2.7 $ }

\newcommand{\pminc}{$17.2$}
\newcommand{\pmaxc}{$189.1$}
\newcommand{\pmind}{$25.1 $}
\newcommand{\pmaxd}{$175.6 $}

\begin{document} 
   \title{Uncovering the true periods of the young sub-Neptunes orbiting TOI-2076}
%\textbf{Author list TBD \& pending Cheops GTO approval} \and
%%%%%%%%%%%%%%%%%%%%%%%%%%%%%%%%%%%%%%%%%%%%%%%%%%%%%%%
%.    WARNING:  DO NOT EDIT YOUR AUTHOR INFO BELOW    %
%%%%%%%%%%%%%%%%%%%%%%%%%%%%%%%%%%%%%%%%%%%%%%%%%%%%%%%
%
% Author info is auto-generated from this google spreadsheet: https://docs.google.com/spreadsheets/d/17KRQMYalFKKTm0ZJO5wHxUhVfgf1chrhw8q7WHVYig0/edit?usp=sharing 
\author{H.P.~Osborn\inst{1}\fnmsep\inst{2}\textsuperscript{,\orcidicon{0000-0002-4047-4724}}\thanks{E-mail:hugh.osborn@space.unibe.ch} \and % hugh.osborn@space.unibe.ch First author
A.~Bonfanti\inst{3}\orcidicon{0000-0002-1916-5935} \and % andrea.bonfanti@oeaw.ac.at Stellar parameters
D.~Gandolfi\inst{4}\orcidicon{0000-0001-8627-9628} \and % davide.gandolfi@unito.it Advisor to Cheops duotransits team & Explore Program manager
C.~Hedges\inst{5}\fnmsep\inst{6}\orcidicon{0000-0002-3385-8391} \and % christina.l.hedges@nasa.gov TESS lightcurve, knowledge of duotransits system
A.~Leleu\inst{7}\orcidicon{0000-0003-2051-7974} \and % Adrien.Leleu@unige.ch MMR analysis
A.~Fortier\inst{1}\fnmsep\inst{8}\orcidicon{0000-0001-8450-3374} \and % andrea.fortier@unibe.ch Cheops GTO team; Cheops Science Enabler nom. by DQ
D.~Futyan\inst{7} \and % David.Futyan@unige.ch Cheops Science Enabler nom. by DQ
P.~Gutermann\inst{9}\fnmsep\inst{10} \and % pascal.guterman@lam.fr Cheops Science Enabler nom. by DQ
P.~F.~L.~Maxted\inst{11} \and % p.maxted@keele.ac.uk Cheops GTO team; Cheops Science Enabler nom. by DQ
L.~Borsato\inst{12}\orcidicon{0000-0003-0066-9268} \and % luca.borsato@inaf.it Stability analysis
K.A.~Collins\inst{13}\orcidicon{0000-0001-6588-9574} \and % karen.collins@cfa.harvard.edu SG1 coordination and data analysis
J.~Gomes~da~Silva\inst{14}\orcidicon{0000-0001-8056-9202} \and % Joao.Silva@astro.up.pt TS3 analysis
Y.~Gómez~Maqueo~Chew\inst{15}\orcidicon{0000-0002-7486-6726} \and % ygmc@astro.unam.mx SAINT-EX (major)
M.~J.~Hooton\inst{1}\orcidicon{0000-0003-0030-332X} \and % matthew.hooton@unibe.ch Support with ground-based data
M.~Lendl\inst{7}\orcidicon{0000-0001-9699-1459} \and % monika.lendl@oeaw.ac.at Advisor to Cheops duotransits team
H.~Parviainen\inst{16}\fnmsep\inst{17}\orcidicon{0000-0001-5519-1391} \and % hannu@iac.es MuSCAT-3 extraction
S.~Salmon\inst{7} \and % sebastien.salmon@uliege.be TS3 analysis
N.~Schanche\inst{8}\orcidicon{0000-0002-9526-3780} \and % nicole.schanche@unibe.ch SAINT-EX (major)
L.M.~Serrano\inst{4} \and % luisamaria.serrano@unito.it Explore involvement. Paper comments
S.G.~Sousa\inst{14}\orcidicon{0000-0001-9047-2965} \and % sergio.sousa@astro.up.pt Stellar parameters
A.~Tuson\inst{18}\orcidicon{0000-0002-2830-9064} \and % alt59@cam.ac.uk Key member of Cheops duotransits team
S.~Ulmer-Moll\inst{7}\orcidicon{0000-0003-2417-7006} \and % solene.ulmer-moll@unige.ch Key member of Cheops duotransits team
V.~Van~Grootel\inst{19}\orcidicon{0000-0003-2144-4316} \and % valerie.vangrootel@uliege.be Stellar parameters
R.D.~Wells\inst{8}\orcidicon{0000-0002-7240-8473} \and % robert.wells@unibe.ch SAINT-EX (major)
T.G.~Wilson\inst{20}\orcidicon{0000-0001-8749-1962} \and % tgw1@st-andrews.ac.uk Stellar parameters
Y.~Alibert\inst{1}\orcidicon{0000-0002-4644-8818} \and % yann.alibert@unibe.ch Cheops GTO team
R.~Alonso\inst{21}\fnmsep\inst{22}\orcidicon{0000-0001-8462-8126} \and % ras@iac.es Cheops GTO team
G.~Anglada\inst{23}\fnmsep\inst{24} \and % anglada@ice.csic.es Cheops GTO team
J.~Asquier\inst{25} \and % joel.asquier@esa.int  CHEOPS Technical Staff
D.~Barrado~y~Navascues\inst{26} \and % barrado@cab.inta-csic.es Cheops GTO team
W.~Baumjohann\inst{27}\orcidicon{0000-0001-6271-0110} \and % baumjohann@oeaw.ac.at Cheops GTO team
T.~Beck\inst{1} \and % thomas.beck@csh.unibe.ch Cheops GTO team
W.~Benz\inst{1}\fnmsep\inst{8} \and % willy.benz@unibe.ch Cheops GTO team
F.~Biondi\inst{28}\fnmsep\inst{29} \and % biondi@mpe.mpg.de;federico.biondi@inaf.it CHEOPS Technical Staff
X.~Bonfils\inst{30} \and % xavier.bonfils@obs.ujf-grenoble.fr Cheops GTO team
F.~Bouchy\inst{7} \and % Francois.Bouchy@unige.ch SAINT-EX
A.~Brandeker\inst{31}\orcidicon{0000-0002-7201-7536} \and % alexis@astro.su.se Cheops GTO team
C.~Broeg\inst{1}\fnmsep\inst{8}\orcidicon{0000-0001-5132-2614} \and % christopher.broeg@unibe.ch Cheops GTO team
T.~Bárczy\inst{32} \and % tamas.barczy@admatis.com Cheops GTO team
S.C.C.~Barros\inst{14}\fnmsep\inst{33}\orcidicon{0000-0003-2434-3625} \and % susana.barros@astro.up.pt Cheops GTO team
J.~Cabrera\inst{34}\orcidicon{0000-0001-6653-5487 } \and % juan.cabrera@dlr.de Cheops GTO team
S.~Charnoz\inst{35} \and % charnoz@ipgp.fr Cheops GTO team
A.~Collier~Cameron\inst{20}\orcidicon{0000-0002-8863-7828} \and % acc4@st-andrews.ac.uk Cheops GTO team
S.~Csizmadia\inst{34} \and % szilard.csizmadia@dlr.de Cheops GTO team
M.~B.~Davies\inst{36}\orcidicon{0000-0001-6080-1190} \and % melvyn_b.davies@math.lu.se Cheops GTO team
M.~Deleuil\inst{9} \and % magali.deleuil@lam.fr Cheops GTO team
L.~Delrez\inst{37}\fnmsep\inst{38} \and % ldelrez@uliege.be Cheops GTO team
B.-O.~Demory\inst{39} \and % brice.demory@csh.unibe.ch Cheops GTO team; Saint-Ex contribution
D.~Ehrenreich\inst{7} \and % david.ehrenreich@unige.ch Cheops GTO team
A.~Erikson\inst{34} \and % Anders.Erikson@dlr.de Cheops GTO team
L.~Fossati\inst{27} \and % Luca.Fossati@oeaw.ac.at Cheops GTO team
M.~Fridlund\inst{40}\fnmsep\inst{41} \and % fridlund@strw.leidenuniv.nl Cheops GTO team
M.~Gillon\inst{37} \and % michael.gillon@ulg.ac.be Cheops GTO team
M.A.~Gómez-Muñoz\inst{42}\orcidicon{0000-0002-3938-4211} \and % mgomez@astro.unam.mx SAINT-EX
M.~Güdel\inst{43} \and % manuel.guedel@univie.ac.at Cheops GTO team
K.~Heng\inst{39}\fnmsep\inst{44} \and % kevin.heng@csh.unibe.ch Cheops GTO team
S.~Hoyer\inst{9} \and % sergio.hoyer@lam.fr Cheops GTO team
K.~G.~Isaak\inst{45} \and % Cheops GTO team
L.~Kiss\inst{46}\fnmsep\inst{47} \and % kiss@konkoly.hu Cheops GTO team
J.~Laskar\inst{48} \and % laskar@imcce.fr Cheops GTO team
A.~Lecavelier~des~Etangs\inst{49} \and % lecaveli@iap.fr Cheops GTO team
C.~Lovis\inst{7} \and % christophe.lovis@unige.ch Cheops GTO team
D.~Magrin\inst{12} \and % demetrio.magrin@inaf.it Cheops GTO team
L.~Malavolta\inst{50}\fnmsep\inst{12}\orcidicon{0000-0002-6492-2085} \and % luca.malavolta@unipd.it Contribution of GAPS data to TS3 analysis
J.~McCormac\inst{44} \and % j.j.mccormac@warwick.ac.uk SAINT-EX
V.~Nascimbeni\inst{12} \and % valerio.nascimbeni@inaf.it Cheops GTO team
G.~Olofsson\inst{31} \and % olofsson@astro.su.se Cheops GTO team
R.~Ottensamer\inst{43} \and % roland.ottensamer@univie.ac.at Cheops GTO team
I.~Pagano\inst{51} \and % isabella.pagano@inaf.it Cheops GTO team
E.~Pallé\inst{21}\fnmsep\inst{52}\orcidicon{0000-0003-0987-1593} \and % epalle@iac.es Cheops GTO team
G.~Peter\inst{34} \and % gisbert.peter@dlr.de Cheops GTO team
D.~Piazza\inst{1} \and % daniele.piazza@unibe.ch CHEOPS Technical Staff
G.~Piotto\inst{50}\fnmsep\inst{12} \and % giampaolo.piotto@unipd.it Cheops GTO team
D.~Pollacco\inst{44} \and % d.pollacco@warwick.ac.uk Cheops GTO team
D.~Queloz\inst{7}\fnmsep\inst{53} \and % didier.queloz@unige.ch Cheops GTO team
R.~Ragazzoni\inst{12}\fnmsep\inst{50} \and % roberto.ragazzoni@inaf.it Cheops GTO team
N.~Rando\inst{25} \and % nicola.rando@esa.int Cheops GTO team
H.~Rauer\inst{34}\fnmsep\inst{54} \and % Heike.Rauer@dlr.de Cheops GTO team
C.~Reimers\inst{43} \and % christian.reimers@univie.ac.at CHEOPS Technical Staff
I.~Ribas\inst{23}\fnmsep\inst{24} \and % iribas@ice.cat Cheops GTO team
O.~D.~S.~Demangeon\inst{14}\fnmsep\inst{33}\orcidicon{0000-0001-7918-0355} \and % olivier.demangeon@astro.up.pt Cheops GTO team
A.~M.~S.~Smith\inst{34}\orcidicon{0000-0002-2386-4341} \and % alexis.smith@dlr.de Cheops GTO team
L.~Sabin\inst{42}\orcidicon{0000-0003-0242-0044} \and % lsabin@astro.unam.mx SAINT-EX
N.~Santos\inst{14}\fnmsep\inst{33} \and % nuno.santos@astro.up.pt Cheops GTO team
G.~Scandariato\inst{51}\orcidicon{0000-0003-2029-0626} \and % gaetano.scandariato@inaf.it Cheops GTO team
U.~Schroffenegger\inst{8} \and % urs.schroffenegger@unibe.ch SAINT-EX
R.P.~Schwarz\inst{55}\orcidicon{0000-0001-8227-1020} \and % rpschwarz@comcast.net Contributed to SG1 LCO ground-based photometry
A.~Shporer\inst{2}\orcidicon{0000-0002-1836-3120} \and % shporer@mit.edu Contributed to SG1 LCO ground-based photometry
A.~E.~Simon\inst{1}\orcidicon{0000-0001-9773-2600} \and % attila.simon@unibe.ch Cheops GTO team
M.~Steller\inst{27} \and % Manfred.Steller@oeaw.ac.at Cheops GTO team
G.~M.~Szabó\inst{56}\fnmsep\inst{57} \and % szgy@konkoly.hu; szgy@gothard.hu Cheops GTO team
D.~Ségransan\inst{7} \and % damien.segransan@unige.ch Cheops GTO team
N.~Thomas\inst{1} \and % nicolas.thomas@unibe.ch Cheops GTO team
S.~Udry\inst{7}\orcidicon{0000-0001-7576-6236} \and % stephane.udry@unige.ch Cheops GTO team
I.~Walter\inst{58} \and % Ingo.walter@dlr.de CHEOPS Technical Staff
N.~Walton\inst{59}  % naw@ast.cam.ac.uk Cheops GTO team
}

%%%%%%%%%%%%%%%%%%%%%%%%%%%%%%%%%%%%%%%%%%%%%%%%%%%%%%%
%.    WARNING:  DO NOT EDIT YOUR AUTHOR INFO HERE.    %
%%%%%%%%%%%%%%%%%%%%%%%%%%%%%%%%%%%%%%%%%%%%%%%%%%%%%%%
%
% Author info is auto-generated from this google spreadsheet: https://docs.google.com/spreadsheets/d/17KRQMYalFKKTm0ZJO5wHxUhVfgf1chrhw8q7WHVYig0/edit?usp=sharing 
%
\institute{Physikalisches Institut, University of Bern, Gesellsschaftstrasse 6, 3012 Bern, Switzerland \and
Department of Physics and Kavli Institute for Astrophysics and Space Research, Massachusetts Institute of Technology, Cambridge, MA 02139, USA \and
Austrian Academy of Sciences, Schmiedlstrasse 6, 8042 Graz, Austria \and
Dipartimento di Fisica, Universita degli Studi di Torino, via Pietro Giuria 1, I-10125, Torino, Italy \and
Bay Area Environmental Research Institute, P.O. Box 25, Moffett Field, CA 94035, USA \and
NASA Ames Research Center, Moffett Field, CA, 94035, USA \and
Observatoire Astronomique de l'Université de Genève, Chemin Pegasi 51, Versoix, Switzerland \and
Center for Space and Habitability, University of Bern, Gesellschaftsstrasse 6, 3012, Bern, Switzerland \and
Aix Marseille Univ, CNRS, CNES, LAM, 38 rue Frédéric Joliot-Curie, 13388 Marseille, France \and
Division Technique INSU, CS20330, 83507 La Seyne sur Mer cedex, France \and
Astrophysics Group, Keele University, Staffordshire, ST5 5BG, United Kingdom \and
INAF, Osservatorio Astronomico di Padova, Vicolo dell'Osservatorio 5, 35122 Padova, Italy \and
Center for Astrophysics \textbar \ Harvard \& Smithsonian, 60 Garden Street, Cambridge, MA 02138, USA \and
Instituto de Astrofisica e Ciencias do Espaco, Universidade do Porto, CAUP, Rua das Estrelas, 4150-762 Porto, Portugal \and
Universidad Nacional Aut\'onoma de M\'exico, Instituto de Astronom\'ia, AP 70-264, CDMX  04510, M\'exico \and
Instituto de Astrof\'isica de Canarias, 38200 La Laguna, Tenerife, Spain \and
Dept. Astrof\'isica, Universidad de La Laguna (ULL), E-38206 La Laguna, Tenerife, Spain \and
Astrophysics Group, Cavendish Laboratory, University of Cambridge, J.J. Thomson Avenue, Cambridge CB3 0HE, UK \and
Space sciences, Technologies and Astrophysics Research (STAR) Institute, Universit\'e de Li\`ege, 19C All\'ee du 6 Ao\^ut, B-4000 Li\`ege, Belgium \and
Centre for Exoplanet Science, SUPA School of Physics and Astronomy, University of St Andrews, North Haugh, St Andrews KY16 9SS, UK \and
Instituto de Astrofisica de Canarias, 38200 La Laguna, Tenerife, Spain \and
Departamento de Astrofisica, Universidad de La Laguna, 38206 La Laguna, Tenerife, Spain \and
Institut de Ciencies de l'Espai (ICE, CSIC), Campus UAB, Can Magrans s/n, 08193 Bellaterra, Spain \and
Institut d'Estudis Espacials de Catalunya (IEEC), 08034 Barcelona, Spain \and
ESTEC, European Space Agency, 2201AZ, Noordwijk, NL \and
Depto. de Astrofisica, Centro de Astrobiologia (CSIC-INTA), ESAC campus, 28692 Villanueva de la Cañada (Madrid), Spain \and
Space Research Institute, Austrian Academy of Sciences, Schmiedlstrasse 6, A-8042 Graz, Austria \and
Max Planck Institut für Extraterrestrische Physik. Gießenbachstraße 1, 85748 Garching bei München, Germany \and
INAF - Osservatorio Astronomico di Padova. Vicolo Osservatorio 5, 35122 PADOVA, Italy \and
Université Grenoble Alpes, CNRS, IPAG, 38000 Grenoble, France \and
Department of Astronomy, Stockholm University, AlbaNova University Center, 10691 Stockholm, Sweden \and
Admatis, 5. Kandó Kálmán Street, 3534 Miskolc, Hungary \and
Departamento de Fisica e Astronomia, Faculdade de Ciencias, Universidade do Porto, Rua do Campo Alegre, 4169-007 Porto, Portugal \and
Institute of Planetary Research, German Aerospace Center (DLR), Rutherfordstrasse 2, 12489 Berlin, Germany \and
Université de Paris, Institut de physique du globe de Paris, CNRS, F-75005 Paris, France \and
Centre for Mathematical Sciences, Lund University, Box 118, 22100 Lund, Sweden \and
Astrobiology Research Unit, Université de Liège, Allée du 6 Août 19C, B-4000 Liège, Belgium \and
Space sciences, Technologies and Astrophysics Research (STAR) Institute, Université de Liège, Allée du 6 Août 19C, 4000 Liège, Belgium \and
Center for Space and Habitability, Gesellsschaftstrasse 6, 3012 Bern, Switzerland \and
Leiden Observatory, University of Leiden, PO Box 9513, 2300 RA Leiden, The Netherlands \and
Department of Space, Earth and Environment, Chalmers University of Technology, Onsala Space Observatory, 43992 Onsala, Sweden \and
Universidad Nacional Aut\'onoma de M\'exico, Instituto de Astronom\'ia, AP 106, Ensenada 22800, BC, M\'exico \and
Department of Astrophysics, University of Vienna, Tuerkenschanzstrasse 17, 1180 Vienna, Austria \and
Department of Physics, University of Warwick, Gibbet Hill Road, Coventry CV4 7AL, United Kingdom \and
Science and Operations Department - Science Division (SCI-SC), Directorate of Science, European Space Agency (ESA), European Space Research and Technology Centre (ESTEC), Keplerlaan 1, 2201-AZ Noordwijk, The Netherlands \and
Konkoly Observatory, Research Centre for Astronomy and Earth Sciences, 1121 Budapest, Konkoly Thege Miklós út 15-17, Hungary \and
ELTE E\"otv\"os Lor\'and University, Institute of Physics, P\'azm\'any P\'eter s\'et\'any 1/A, 1117 Budapest, Hungary \and
IMCCE, UMR8028 CNRS, Observatoire de Paris, PSL Univ., Sorbonne Univ., 77 av. Denfert-Rochereau, 75014 Paris, France \and
Institut d'astrophysique de Paris, UMR7095 CNRS, Université Pierre \& Marie Curie, 98bis blvd. Arago, 75014 Paris, France \and
Dipartimento di Fisica e Astronomia "Galileo Galilei", Universita degli Studi di Padova, Vicolo dell'Osservatorio 3, 35122 Padova, Italy \and
INAF, Osservatorio Astrofisico di Catania, Via S. Sofia 78, 95123 Catania, Italy \and
Deptartamento de Astrofisica, Universidad de La Laguna (ULL), E-38206 La Laguna, Tenerife, Spain \and
Cavendish Laboratory, JJ Thomson Avenue, Cambridge CB3 0HE, UK \and
Center for Astronomy and Astrophysics, Technical University Berlin, Hardenberstrasse 36, 10623 Berlin, Germany \and
Patashnick Voorheesville Observatory, Voorheesville, NY 12186, USA \and
ELTE Eötvös Loránd University, Gothard Astrophysical Observatory, 9700 Szombathely, Szent Imre h. u. 112, Hungary \and
MTA-ELTE Exoplanet Research Group, 9700 Szombathely, Szent Imre h. u. 112, Hungary \and
Institute of Optical Sensor Systems, German Aerospace Center (DLR), Rutherfordstra{\ss}e 2, 12489, Berlin, Germany \and
Institute of Astronomy, University of Cambridge, Madingley Road, Cambridge, CB3 0HA, United Kingdom 
}
%%%%%%%%%%%%%%%%%%%%%%%%%%%%%%%%%%%%%%%%%%%%%%%%%%%%%%%
%.    WARNING:  DO NOT EDIT YOUR AUTHOR INFO ABOVE    %
%%%%%%%%%%%%%%%%%%%%%%%%%%%%%%%%%%%%%%%%%%%%%%%%%%%%%%%
%
% Author info is auto-generated from this google spreadsheet: https://docs.google.com/spreadsheets/d/17KRQMYalFKKTm0ZJO5wHxUhVfgf1chrhw8q7WHVYig0/edit?usp=sharing 
%

   \date{Received January 10, 2022; accepted March 1, 2022}

  \abstract
   {TOI-2076 is a transiting three-planet system of sub-Neptunes orbiting a bright (G = 8.9 mag), young ($340\pm80$\,Myr) K-type star. Although a validated planetary system, the orbits of the two outer planets were unconstrained as only two non-consecutive transits were seen in \tess{} photometry. This left 11 and 7 possible period aliases for each.}
   {To reveal the true orbits of these two long-period planets, precise photometry targeted on the highest-probability period aliases is required. Long-term monitoring of transits in multi-planet systems can also help constrain planetary masses through TTV measurements.}
   {We used the \texttt{MonoTools} package to determine which aliases to follow, and then performed space-based and ground-based photometric follow-up of TOI-2076~c and d with \cheops{}, SAINT-EX, and LCO telescopes.}
   {\cheops{} observations revealed a clear detection for TOI-2076~c at $P=$\Tttvperc{}d, and allowed us to rule out three of the most likely period aliases for TOI-2076~d. Ground-based photometry further enabled us to rule out remaining aliases and confirm the $P=$\Tttvperd{}d alias. These observations also improved the radius precision of all three sub-Neptunes to \Trplzero{}, \Trplone{}, and \Trpltwo{}\,$R_\oplus$. Our observations also revealed a clear anti-correlated TTV signal between planets b and c \refcom{likely} caused by their proximity to the 2:1 resonance, while planets c and d appear close to a 5:3 \refcom{period commensurability, although model degeneracy meant we were unable to retrieve robust TTV masses. Their inflated radii, likely due to extended H-He atmospheres, combined with} low insolation makes all three planets excellent candidates for future comparative transmission spectroscopy with JWST.}
   {}

   \maketitle

\section{Introduction}
NASA's Transiting Exoplanet Survey Satellite \citep[\tess;][]{2015JATIS...1a4003R} has excelled in detecting transiting planets around bright stars \citep[e.g.][]{2018ApJ...868L..39H,2019ApJ...875L...7D,2020AJ....160...96T,2020MNRAS.491.2982E,2020AJ....160..129K,2021A&A...648A..75S} and around young stars \citep[e.g.][]{2019ApJ...880L..17N,2019A&A...630A..81B,2020Natur.582..497P,2020AJ....160...33R,2021AJ....161...65N}. Bright transiting planets are amenable to detailed characterisation, including through transmission spectroscopy, while young planets give insights into planetary formation and evolution.
%With radii of \Trplzeroshort{}, \Trploneshort{} and \Trpltwoshort{}$R_\oplus$, this bright multi-planet system... 

However, due to the short 27-d duration of its sectors, \tess\ can struggle with long-period planets with $P$\,>\,15\,d, especially at low ecliptic latitudes where \tess\ sky coverage has thus far been lower. One clear example of this is for planetary candidates seen to transit in two non-consecutive sectors --- the so-called `duotransit' cases. As such, there exists a large array of potential period aliases for each planet, which are compatible with the observed data. This set of period aliases $P \in (t_{{\rm tr},2}-t_{{\rm tr},1})/\{1,2,3, \cdots, N_{\rm max}\}$ are bounded at the long end by the temporal distance between the transits $P_{\rm max}=(t_2 - t_1)$ and at the short end by the non-detection of subsequent transits in the \tess\ data. Such cases are expected to be commonplace during the \tess\ extended mission, as planets that were observed to transit once in the primary mission transit again \citep{2020MNRAS.494..736C,2021MNRAS.500.5088C}.

Without knowledge of an exoplanet's orbit, variables such as the planetary equilibrium temperature are unconstrained, and scheduling future characterisation efforts, such as Rossiter-McLaughlin (RM) measurements or transmission spectroscopy, are difficult or even impossible. Using radial velocity observations to measure a planetary mass is also significantly easier when the orbital period is known a priori from transit photometry, especially for active young stars. For all of these reasons, it is imperative for us to recover the true period of such planets.

The follow-up of such ``Duotransits'' in order to find the correct period is not a new concept. \ktwo\ provided multiple such cases, as was explored by \citet{2020AJ....159...93D}. Two of the planets found by \ktwo\ to orbit HIP\,41378 are duotransiters \citep{2019AJ....157...19B}, and a combination of radial velocities and ground-based transit photometry were able to recover the true period of HIP-41378\,f \citep{2019arXiv191107355S,2021MNRAS.504L..45B}. In \tess, the true period of TOI-2257\,b, which produced two 0.4\% transits in \tess\ Year-2 photometry consistent with four possible period aliases, was recovered through ground-based photometry \citep{2021arXiv211101749S}. However, the majority of the planets so far followed up on in this way typically either show few period aliases, or they produce deep eclipses easily observable from the ground (depth\,$>$\,0.4\%). The most interesting planets -- small planets around bright stars -- are therefore more challenging to observe and solve.

ESA's CHaracterising ExOPlanets Satellite (\cheops) space telescope, which launched in 2019 with a goal of detecting and characterising the transits of small exoplanets \citep{2021ExA....51..109B}, is well placed to perform this search. With a 30\,cm aperture, it can achieve photometric precision of the order of $\sim$15\,ppm over a 6 hour window for a G\,=\,9 mag star. This provides a higher per-transit signal-to-noise ratio (S/N) than \tess, and as such it has been successful in observing and confirming the transiting nature of small, long-period transiting planets including the $P=20.7$\,d, $2.9R_\oplus$ TOI-178\,g \citep{2021A&A...649A..26L}; the $P=29.5$\,d, $2.0\,R_\oplus$ HD\,108236\,f \citep{2021A&A...646A.157B}; and the $P=110$\,d, $2.56\,R_\oplus$ $\nu^2$ Lupi\,d \citep{2021NatAs...5..775D}.

TESS Object of Interest TOI-2076 (TIC27491137) is a system of three transiting sub-Neptunes validated by \citet{2021AJ....162...54H} (hereafter H21).
Orbiting a $\sim$200\,Myr old G\,=\,8.9 mag K-type star, TOI-2076 is both bright and young making it a highly valuable multi-planet system. It initially became a \tess\ object of interest after observations in Sectors 16 \& 23 \citep{2021ApJS..254...39G}, and the photometry revealed a total of only 9 transits - five from the inner \Tttvperbshort d planet, and two each from the planets c \& d (one transit in each of the two sectors), making them both ``Duotransits''. The transits were compatible with 11 possible period aliases for TOI-2076~c between \pminc d and \pmaxc d, and seven aliases for TOI-2076~d between \pmind d and \pmaxd d (as shown in H21). With transit depths of $\lesssim2$\,ppt only space-based photometry, for example with \cheops{}, is able to confidently re-detect the transits of these sub-Neptunes.
%, and meaning even the order of the two planets is, from the TESS observations alone, uncertain.
% such as TOI-2076 where the planetary reflex motion must be disentangled from the stellar noise.
%For TOI-2076c \& d, the two transits for each of the two outer planets produce 

In this paper we detail \cheops{} \& ground-based observations of TOI-2076 which are able to recover the true periods of these two long-period long planets. Section 2 presents the follow-up data, which was obtained on this star, as well as its immediate reduction. Section 3 details the analyses performed with this data, including both the pre- and post-observation analyses. In Section 4 we detail the results of these analyses, and put them in context of the state-of-the art.

%--------------------------------------------------------------------
\section{Data}

\subsection{\tess\ Observations}

\tess\ observed TOI-2076 in sectors 16 and 23 in 2-minute cadence. We use the \tess\ light curves created by H21 to supplement the \cheops\ data, \refcom{which are explained in more detail in H21, Section 2.1}. These light curves use the target pixel file (TPF) products from the SPOC pipeline from sectors 16 and 23. Cadences with significantly poor data quality are removed. Light curves are built taking the pipeline aperture, and detrended using \texttt{lightkurve}'s \texttt{RegressionCorrector} tool \citep{cardoso2018lightkurve}. The final composite light curve is detrended against a linear combination of i) significant trends of pixels outside the aperture, ii) the mean and standard deviation of the mission quaternions, and iii) a b-spline. Together these components remove scattered light background, jitter and stellar variability, respectively. Cadences expected to contain transits were masked in this fit. This produces a light curve which has improved precision over the pipeline products, \refcom{as can be seen in H21, Figures 1 \& 2}.

\subsection{\cheops\ Observations}

Through the \cheops\ 
Guaranteed Time Observations (GTO) programme CH\_PR110048 ("Duos - Recovering long period duo-transiting planets"), we scheduled multiple observations of period aliases for TOI-2076 c \& d.
The observing strategy was dictated by determining the marginal probability for each alias (described in Section~\ref{sect:premod}) and observing aliases with $p$\,>\,2\%. The strategy was then adapted for each new observation we received. In total this led to a single visit of a TOI-2076\,c period alias, and two visits of TOI-2076\,d period aliases. We also re-observed a transit of the inner planet, TOI-2076\,b, to improve radius precision and potentially detect transit timing variations (TTVs).

The \cheops\ data were processed by the most recent Data Reduction pipeline DR13 \citep{2020A&A...635A..24H}. We downloaded \cheops{} data from \texttt{DACE} \citep{2015ASPC..495....7B} using the \texttt{pycheops} interface \citep{maxted2021analysis}, and chose the decontaminated \texttt{OPTIMAL} light curve. We then clipped outliers, using both the in-built \texttt{pycheops} default function, and then a further step to clip any points with a background value larger than 0.2, or a flux outside of the range of $-5<({\rm flux/ppt})<5$. We also extracted important decorrelation parameters including centroid position, background, roll-angle, smear, etc.
\refcom{The raw \& detrended \cheops{} data presented here is available through CDS.}

Scheduling continuous transit observations at high-efficiency is a complex problem and, due to competition between the many targets \& programmes on \cheops, not all planned observations can typically be observed. This meant \cheops\ did not cover all high-probability period aliases for TOI-2076\,d and was unable to recover a period, leaving possible aliases at 25.1 and 35.1\,d. Therefore, in order to confirm the orbital period, we turned to ground-based observatories.

\begin{table*}
    \centering
    \caption{Key information for all of the photometry presented in this paper.}
    \label{tab:photometry}
    \tiny{
    \begin{tabular}{lcccccccc}
        \hline
        \hline
        -- & Start time [UT] & Start time [BJD] & Dur [hrs] & Exp [s] & cad [s] & pl. & aliases [d] & File ref. \\
        \hline
Cheops visit 1 & 2021-02-28 09:02:04 & 2459273.87644  &  8.884 & 42.0  & 42.0 &  c  &  21.014d  &  CH\_PR110048\_TG002501\_V0200 \\
Cheops visit 2 & 2021-04-28 18:41:46 & 2459333.27901  &  10.553 & 42.0  & 42.0 &  d  &  43.907d  &  CH\_PR110048\_TG003201\_V0200 \\
Cheops visit 3 & 2021-04-29 07:13:25 & 2459333.80099  &  9.771 & 42.0  & 42.0 &  b  &  10.355d  &  CH\_PR110048\_TG003601\_V0200 \\
LCO/Sinistro (z') & 2021-05-05 02:40:12 & 2459339.61126  &  5.804 & 36.0  & 44.9 &  d  &  25.090d  &  -- \\
Cheops visit 4 & 2021-05-13 10:17:08 & 2459347.92857  &  10.168 & 42.0  & 42.0 &  d  &  29.3 \& 58.5d  &  CH\_PR110048\_TG003701\_V0200 \\
Saint-Ex (r') & 2021-05-25 03:17:19 & 2459359.63704  &  3.21 & 8.0  & 23.7 &  d  &  35.125d  &  -- \\
LCO/MuSCAT3 (g') & 2021-06-29 06:07:49 & 2459394.75544  &  4.75 & 37.0  & 42.3 &  d  &  35.125d  &  -- \\
LCO/MuSCAT3 (r') & 2021-06-29 06:07:41 & 2459394.75534  &  4.75 & 21.0  & 26.1 &  d  &  35.125d  &  -- \\
LCO/MuSCAT3 (i') & 2021-06-29 06:07:36 & 2459394.75528  &  4.756 & 19.0  & 24.1 &  d  &  35.125d  &  -- \\
LCO/MuSCAT3 (z') & 2021-06-29 06:07:43 & 2459394.75536  &  4.749 & 25.0  & 28.1 &  d  &  35.125d  &  -- \\
    \hline
    \end{tabular}
    \begin{tablenotes}
    \item "Dur" refers to the visit duration in hours, "Exp" the exposure time, while "cad" is the cadence (i.e. median gap between subsequent exposures, including overheads), "pl." distinguishes which of the three TOI-2076 planets was targeted. "File ref." refers to the unique file reference key generated by the Cheops DRP.
     \end{tablenotes}
     }
\end{table*}

\subsection{LCO/McDonald observations}

We observed a transit window of the 25.09\,d alias of TOI-2076~d in Pan-STARRS $z$-short band on UTC 2021 May 05 from the Las Cumbres Observatory Global Telescope \citep[LCOGT or LCO;][]{Brown:2013} 1.0\,m network node at McDonald Observatory. We used the {\tt TESS Transit Finder}, which is a customised version of the {\tt Tapir} software package \citep{Jensen:2013}, to schedule our transit observations. The 1\,m class telescopes are equipped with 4096\,$\times$\,4096 pixel SINISTRO cameras having an image scale of $0.389$\arcsec\ per pixel, resulting in a $26\arcmin\times26\arcmin$ field of view.
Exposures were defocused to improve efficiency and photometric precision.
The images were calibrated by the standard LCOGT {\tt BANZAI} pipeline \citep{McCully:2018}, and photometric data were extracted using {\tt AstroImageJ} \citep{Collins:2017}. The images have a typical stellar point-spread-function with a FWHM of $\sim$5$\arcsec$, and circular photometric apertures with radius $7.4$\arcsec\ were used to extract the differential photometry. The target star photometric aperture excludes flux from the nearest Gaia EDR3 neighbours.
\refcomb{Raw and detrended LCO/McDonald photometry is available on CDS.}

\subsection{SAINT-EX observations}
In an effort to catch the 35.1\,d alias of planet d, TOI-2076 was observed on the night of 2021-05-25 between 03:17 and 06:29 UT from the SAINT-EX telescope at the San Pedro M\'{a}rtir observatory, Mexico \citep{Demory2020}. SAINT-EX is a 1-metre F/8 Ritchey-Chretien telescope built to be complementary to the \texttt{SPECULOOS} network of telescopes, which are focused on searching for transiting planets around ultra-cool dwarfs \citep{2018csss.confE..59S, Sebastian2021}.

Due to its 12$\arcmin$ field of view, it was not possible to include any bright comparison stars in the same field as TOI-2076. The observations were made using the r$'$ filter. In order to increase the efficiency and avoid saturation of the bright target, the telescope was defocused, producing a ringed PSF with a diameter of $\sim$10 pixels.

We performed simple image reduction and extracted source counts for TOI-2076 and 7 comparison stars using \texttt{AstroImageJ} (or \texttt{AIJ}), setting an aperture with a radius of 30 pixels, and extracting background flux from an annulus between 47 and 58 pixels in distance from each source.
As well as total fluxes for each star, we also extracted meta-data including airmass, PSF width, PSF FWHM, X \& Y centroids, PSF roundness, which were used to help decorrelate the light curve.
\refcomb{Raw and detrended Saint-Ex photometry is available on CDS.}

\subsection{LCO/MuSCAT observations}
A Director's Discretionary Time proposal on the LCOGT network was also approved to observe and confirm the $P=35.1$\,d alias. An ingress of this alias was visible from Haleakala, Hawaii on 2021-06-29 (BJD=2459394.95), and we scheduled a 4.8\,hr observation of TOI-2076 with the MuSCAT-3 instrument on the 2.0m Faulkes Telescope North \citep{2020SPIE11447E..5KN}. The MuSCAT-3 instrument is able to simultaneously observe in g, r, i \& z filters at different exposure lengths, enabling photometric observations with high efficiency \citep{narita2015muscat}. Due to the bright nature of TOI-2076, we opted to perform the observations with the diffuser in place, thereby allowing longer exposures without assymetric PSFs caused by defocusing. We used exposure times of 37, 19, 21 \& 25 seconds respectively and the \texttt{FAST} read-out mode, resulting in 405, 647, 701, \& 605 exposures respectively.

The small field of view of MuSCAT-3 meant that no similar-brightness stars were present within the field. Extraction was performed using a combination of \texttt{AstroImageJ} and the MuSCAT-3 pipeline \footnote{\url{https://github.com/hpparvi/MuSCAT2_transit_pipeline}.}.

The MuSCAT-3 pipeline produced aperture photometry with less scatter and therefore we used this as our flux input. The ``entropy'' parameter computed by the pipeline (flux inside the photometry aperture normalised by the total aperture flux) was also extracted as a useful detrending parameter. From the \texttt{AIJ} analysis, we extracted the more complete \refcom{meta-data}, including sum of comparison star flux, PSF width, x \& y centroids, etc.
\refcomb{Raw and detrended LCO/MuSCAT-3 photometry is available on CDS.}

%--------------------------------------------------------------------
\section{Analysis}

\subsection{Stellar Parameters}
\subsubsection{Bulk Physical Properties}
\label{sect:starpars}
In order to derive precise stellar parameters, we used spectra taken with HARPS-N at Telescopio Nazionale Galileo, in the framework of the Global Architecture of Planetary Systems (GAPS) project  \citep[see e.g. ][]{2013A&A...554A..28C,2020A&A...638A...5C}. 
64 spectra taken between 2020-08-06 and 2021-06-14 were co-added into a single stacked spectrum which had an average S/N of around 650 at 550~nm.
We then derived the stellar atmospheric parameters ($T_{\mathrm{eff}}$, $\log g$, microturbulence, [Fe/H]), and its respective uncertainties using ARES+MOOG, following the same methodology described in \citet{Santos-13} and \citet{Sousa-14}. We measured the equivalent widths (EW) of iron lines using the \texttt{ARES} code\footnote{The last version of \texttt{ARES} code (\texttt{ARES} v2) can be downloaded at \url{https://github.com/sousasag/ARES}.} \citep{Sousa-07, Sousa-15}. A minimisation process was used to find the ionisation and excitation equilibrium once it converges to the best set of spectroscopic parameters. This process uses a grid of Kurucz model atmospheres \citep{Kurucz-93} and the radiative transfer code \texttt{MOOG} \citep{Sneden-73}. We obtained a temperature of \refcom{5200\,$\pm$\,70\,K}, a log\,g of 4.45\,$\pm$\,0.12 dex, a [Fe/H] of -0.09\,$\pm$\,0.04, and a microturbulance velocity of 1.08\,$\pm$\,0.05\kms{}.

To compute the stellar radius of TOI-2076, we used a modified infrared flux method (IRFM; \citealt{Blackwell1977}) to determine the stellar angular diameter and effective temperature via a Markov-Chain Monte Carlo (MCMC) approach, as recently detailed in \citet{Schanche2020}. As these properties can be derived from the stellar apparent bolometric flux, we produce synthetic photometry by constructing spectral energy distributions (SEDs) from stellar atmospheric models using the stellar parameters derived from our spectral analysis as priors that we attenuate to account for reddening with the extinction left as a free parameter. The computed synthetic fluxes were compared with the retrieved broadband fluxes and uncertainties from the most recent data releases for the following bandpasses; {\it Gaia} G, G$_{\rm BP}$, and G$_{\rm RP}$, 2MASS J, H, and K, and {\it WISE} W1 and W2 \citep{Skrutskie2006,Wright2010,GaiaCollaboration2021}. To include any systematic uncertainties derived from stellar atmospheric model differences in our stellar radius error we used a Bayesian modelling averaging method with stellar models from a range of \textsc{atlas} \citep{Kurucz1993,Castelli2003} catalogues in order to produce weighted averaged posterior distributions. From this analysis we find a $T_{\mathrm{eff}}$ and $E(B-V)$ of 5181$\pm$37\,K and 0.02$\pm$0.01, respectively. Lastly, we converted the stellar angular diameter of TOI-2076 to the radius using the offset corrected {\it Gaia} EDR3 parallax \cite{Lindegren2021}, and obtain a $R_s$=\TRs{}\,$R_{\odot}$.

The set given by ($T_{\mathrm{eff}}$, [Fe/H], $R_s$) is then assumed as input to derive the isochronal mass $M_s$ and age $t_s$. To this end, we used the isochrone placement technique \citep{bonfanti15,bonfanti16} applied to pre-computed grids of PARSEC\footnote{\textit{P}adova \textit{A}nd T\textit{R}ieste \textit{S}tellar \textit{E}volutionary \textit{C}ode: \url{http://stev.oapd.inaf.it/cgi-bin/cmd}} v1.2S \citep{marigo17} isochrones and tracks to compute a first pair of mass and age estimates. Furthermore, we derived a second pair of mass and age values by directly fitting the input set into the evolutionary tracks built by the CLES\footnote{Code Liègeois d'Évolution Stellaire} code \citep{scuflaire08}, following the Levenberg-Marquadt minimisation scheme presented in \citet{salmon21}.
Our adopted $M_s=0.824_{-0.037}^{+0.035}\,M_{\odot}$ and $t_s=4.5_{-3.3}^{+3.1}$ Gyr values are finally computed by merging the two respective pairs of distributions inferred from the two different evolutionary models, after checking their mutual consistency using the $\chi^2$-based criterion described in detail in \citet{2021A&A...646A.157B}.
The derived parameters are in agreement (at the $1\,\sigma$ level) with those derived by H21.

\subsubsection{Stellar Age}
\refcomb{H21 presented multiple lines of evidence for the youth of both TOI-2076 and TOI-1807, a separate transiting planet host close-by and co-moving with TOI-2076 which likely formed together.
This included gyrochronology ($125-230$\,Myr), $\log{\rm R'HK}$ ($12-870$\,Myr), Li absorption ($<800$\,Myr), Ca II IR triplet core emission ($<1000$\,Myr) and X-ray flux ($>18$\,Myr), giving a combined age of $200\pm50$\,Myr.
We chose to re-assess the age given our follow-up spectra and more precise stellar parameters.}

We derived \refcom{the} Mount Wilson Ca II index ($\log{\rm R'HK}$ -- the chromospheric contribution of the H and K Ca lines) from the stacked HARPS-N spectra of $-4.373\pm0.02$ \refcomb{using ACTIN \citep{2018JOSS....3..667G}}.
The relation of \citet{2016A&A...594L...3L} allows us to convert this to a stellar age of $0.42\pm0.13$\,Gyr \refcomb{-- far more precise than that of H21}.
\refcomb{We also re-derived a gyrochronological age} using the relation of \citet{2008ApJ...687.1264M} and the rotation period derived by H21 ($6.84\pm0.58$d), finding a slightly older age of $0.25\pm0.12$ Gyr.
Both these techniques therefore support a young ($<0.5$Gyr)\refcomb{ age, and are in agreement with the independent analyses presented in H21.}
We adopt the weighted mean of the two activity-derived ages as our derived age of TOI-2076 going forward - $0.34\pm 0.08$Gyr.

These stellar ages are at odds with that derived from our isochrones, which \refcomb{is imprecise but suggests an intermediate-age star ($4.5^{+3.1}_{-3.3}$).}
However, isochronal ages are frequently in tension with astroseismology and activity-derived ages \citep[e.g.][]{2005ESASP.576..187P,brown14,kovacs15}, \refcomb{therefore we choose not to include it in our derived average age.}
%Due to this, and the large uncertainties on the isochronal age, 
%\textbf{Update Table 1 adding also $t_s$ and [Fe/H]}

\begin{table}
\caption{Derived stellar parameters.}  % title of Table
\label{table:star_params}      % is used to refer this table in the text
\centering                          % used for centering table
\begin{tabular}{l c }        % centred columns (4 columns)
\hline\hline                 % inserts double horizontal lines
Parameter & Value \\
\hline                        % inserts single horizontal line
Name & TOI-2076 \\
TIC & \TICstar{}$^{\dagger}$ \\
\textit{BD} designation & BD+40 2790 \\
\gaia{} DR2 ID & 1490845442647992960 $^{\star}$ \\
RA [$^\circ$, J2015.5] & 217.391994602 $^{\star}$ \\
Dec [$^\circ$, J2015.5] & 39.790398204 $^{\star}$ \\
\tess{} mag & $8.3745 \pm 0.006$\,$^{\dagger}$\\
\textit{G} mag & $8.92\pm0.000477$\,$^{\star}$ \\
\textit{K} mag & $7.115\pm0.017$\,$^{\ddagger}$\\
\teff{} [K] & $5200\pm70$\,$^{\beta}$\\
$R_s$ [$R_\odot$] & $0.77\pm0.006$\,$^{\beta}$\\
$M_s$ [$M_\odot$] & $0.824_{-0.037}^{+0.035}$\,$^{\beta}$\\
\logg{} [cgs] & $4.45\pm0.12$\,$^{\beta}$\\
$[{\rm Fe}/{\rm H}]$ & $-0.09\pm0.04$\,$^{\beta}$\\
$\log{\rm R'HK}$ [dex] & $-4.373\pm0.02$\,$^{\beta}$ \\
Gyrochron. Age [Gyr] & $0.204\pm0.050$\,$^{\alpha}$\\
$\log{\rm R'HK}$ Age [Gyr] & $0.42\pm0.13$\,$^{\beta}$\\
Adopted Age [Gyr] & $0.34\pm 0.08$\,$^{\beta}$\\
\hline                                   %inserts single line
\end{tabular}
\begin{tablenotes}
\small
\item Notation refers to the following sources: $^{\star}$ Gaia EDR3 \citep{2021A&A...649A...1G}; $^{\dagger}$ TESS Input Catalog  \citep{2018AJ....156..102S}; $^{\ddagger}$ 2MASS  \citep{2003tmc..book.....C}; $^{\alpha}$ analysis by H21; $^{\beta}$ Our own analysis as described in Section \ref{sect:starpars}.
\end{tablenotes}
\end{table}

\subsection{Photometry -- TESS-only analysis}
\label{sect:premod}
In order to determine which aliases to observe, we first performed model fits to the available \tess{} transits. Typically transit modelling relies on a known orbital period in order to constrain not just the orbital parameters, but also those parameters which determine the transit shape, such as the transit duration and impact parameter, which are influenced by orbit through limits on the planetary velocity. In our case, such constraints need to be inverted -- we must use the transit shape to constrain the orbital velocity (and therefore orbital period). With this goal in mind, we developed the \monotools{} package, which is able to model transit lightcurves in cases of multiple transits, duotransits and monotransits, as well as multiple systems with combinations of such candidates, with both radial velocities and transit photometry \footnote{\url{https://github.com/hposborn/MonoTools}.}.

For such fits, impact parameter, transit duration, and radius ratio are fitted together in a way that is agnostic of the exoplanet orbit. The combination of these transit shape parameters, along with a stellar density constrained from stellar parameters, implies a unique transverse planetary velocity. In the inverse case -- where transit shape constrains orbital parameters - this is known as the photoeccentric effect \citep[e.g.][]{2012ApJ...756..122D}. Converting this velocity directly to a single orbital period parameter and trimming the samples to those regions round period aliases would be incredibly inefficient as the vast majority of derived orbital periods would not fall within these discreet period alias ``island''. So instead, \monotools{} calculates a marginalised probability distribution across all allowed aliases for a given transit model by combining priors for each alias.

A major part of this is the period prior of $P^{-8/3}$ as derived by \citet{2018RNAAS...2..223K}. 
This is necessary as short-period orbits are highly favoured over long-period ones due to a combination of geometric probability and window function. Secondly, a prior is calculated using the probability of the implied orbital velocity given some prior eccentricity distribution. Exoplanet population studies show that planets, especially in multi-planet systems, have a general distribution that peaks at low eccentricities. These population-derived distributions \citep[e.g.][]{2013MNRAS.434L..51K,2015ApJ...808..126V} also imply a probability distribution of orbital velocities relative to the velocity of a circular orbit. This is because velocities much faster or much slower than that of a circular orbit are disfavoured as they imply highly eccentric orbits, which exoplanet population studies show are uncommon \citep{2013MNRAS.434L..51K}, especially in short-period (P<100d) multi-planet systems \citep{2015ApJ...808..126V}. Instead of performing  this step analytically (which requires a complex and infeasible integration over the eccentricity prior), \monotools{} uses pre-computed interpolations for the velocity prior calculated numerically.

The boost to geometric transit probability for eccentric orbits, and the effect of a maximum eccentricity are also considered in this interpolated function. 
%In single systems, orbits which graze the star ($(1-e)*a<2R_s$) provide a maximum eccentricity bound.
In the case of a multi-planet system, orbits which graze (i.e. enter the \refcom{Hill} spheres of) interior planets can be rejected and therefore provide a maximum eccentricity. We use a simple 3-part logmass-radius relation derived from fitting observed exoplanets in order to compute \refcom{Hill} spheres on-the-fly.
%from star-grazing orbits with $a(1-e) > 2R_s; e_{\rm max} = 1 - 2R_s/a$
By modelling all planets simultaneously, the inner planet transits can also improve knowledge of the stellar density, hence improving the derived orbital parameters from transit shape. 

\begin{figure}
   \centering
   \includegraphics[width=\columnwidth]{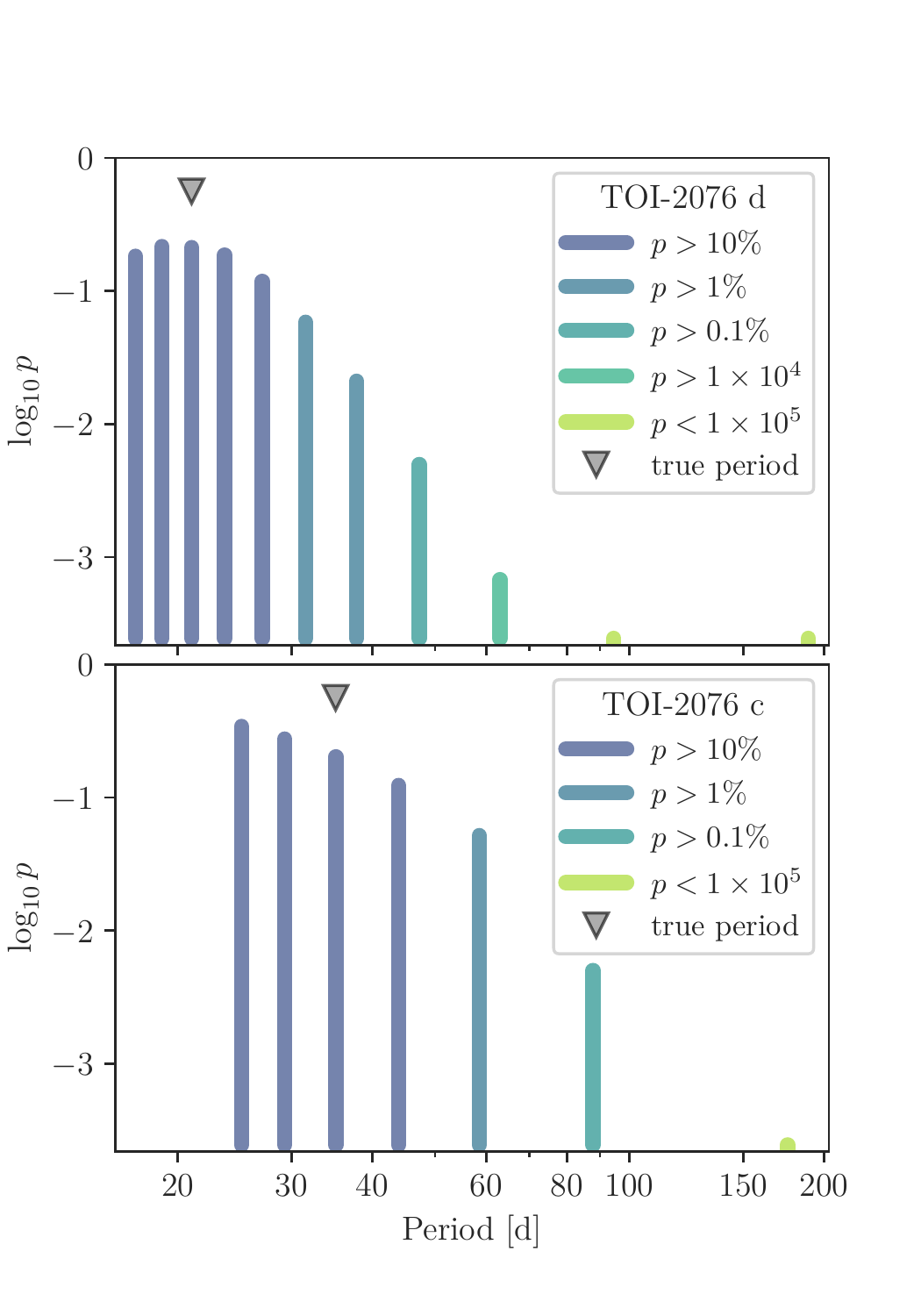}
      \caption{Marginalised $\log_{10}{}$ probabilities for each of TOI-2076~c (upper) and TOI-2076~d (lower) period aliases, as computed by \texttt{MonoTools} before \cheops{} observations.}
         \label{fig:precheopscd}
   \end{figure}

For TOI-2076, we used the eccentricity distribution of \citet{2015ApJ...808..126V}, as this is applicable to short-period transiting multi-planet systems as observed by \tess{}.
We also included a Gaussian Process with a simple harmonic oscillator kernel (\texttt{SHOTerm}) using \celerite{} \citep{celerite1,celerite2} which was pre-trained on out-of-transit data, and has quality factor set to $Q=1/\sqrt{2}$, which is typical for stellar noise.
The resulting posterior probabilities for each period alias are found in Figure \ref{fig:precheopscd}.

We then found the highest-probability aliases which together would give us a $\gtrsim90\%$ probability of a transit redetection.
These were then scheduled on \cheops{}, with the highest-probability aliases of each planet being given highest priority in the \cheops{} scheduler.
This was a total of 5 TOI-2076\,c aliases and 4 TOI-2076\,d aliases.

\begin{figure}
   \centering
   \includegraphics[width=\columnwidth]{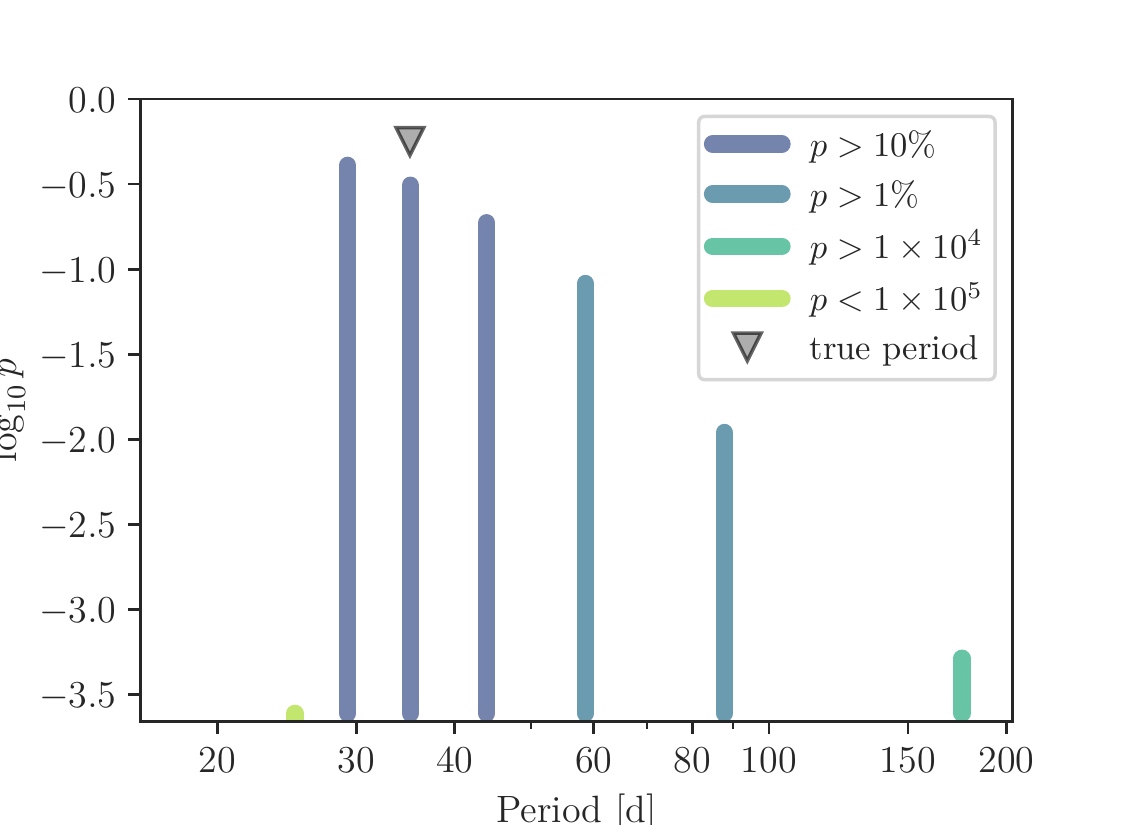}
      \caption{Marginalised $\log_{10}{}$ probabilities for TOI-2076~d period aliases, as computed by \texttt{MonoTools} after the detection of the P=21d alias of TOI-2076~c and before \cheops{} observations of TOI-2076~d. The log probability of the $P=25.1$d alias is far below the y axis limit with $\log_{10}{p}=-19.59$.}
         \label{fig:precheopsd}
   \end{figure}

After the detection of a unique period for TOI-2076\,c, we re-performed this analysis, the resulting marginalised probability distributions are shown in Figure \ref{fig:precheopsd}.
The presence of a planet on a 21\,d orbit interior to planet d drastically reduced the probability of the inner-most alias due to \monotools{} rejecting orbits intersecting with the \refcom{Hill} sphere of TOI-2076\,c.
We updated our \cheops{} observations accordingly, focusing on aliases between 29 and 45\,d.

\subsection{Final combined model}
\label{sect:combmod}
\begin{figure*}
   \centering
   \includegraphics[width=\textwidth]{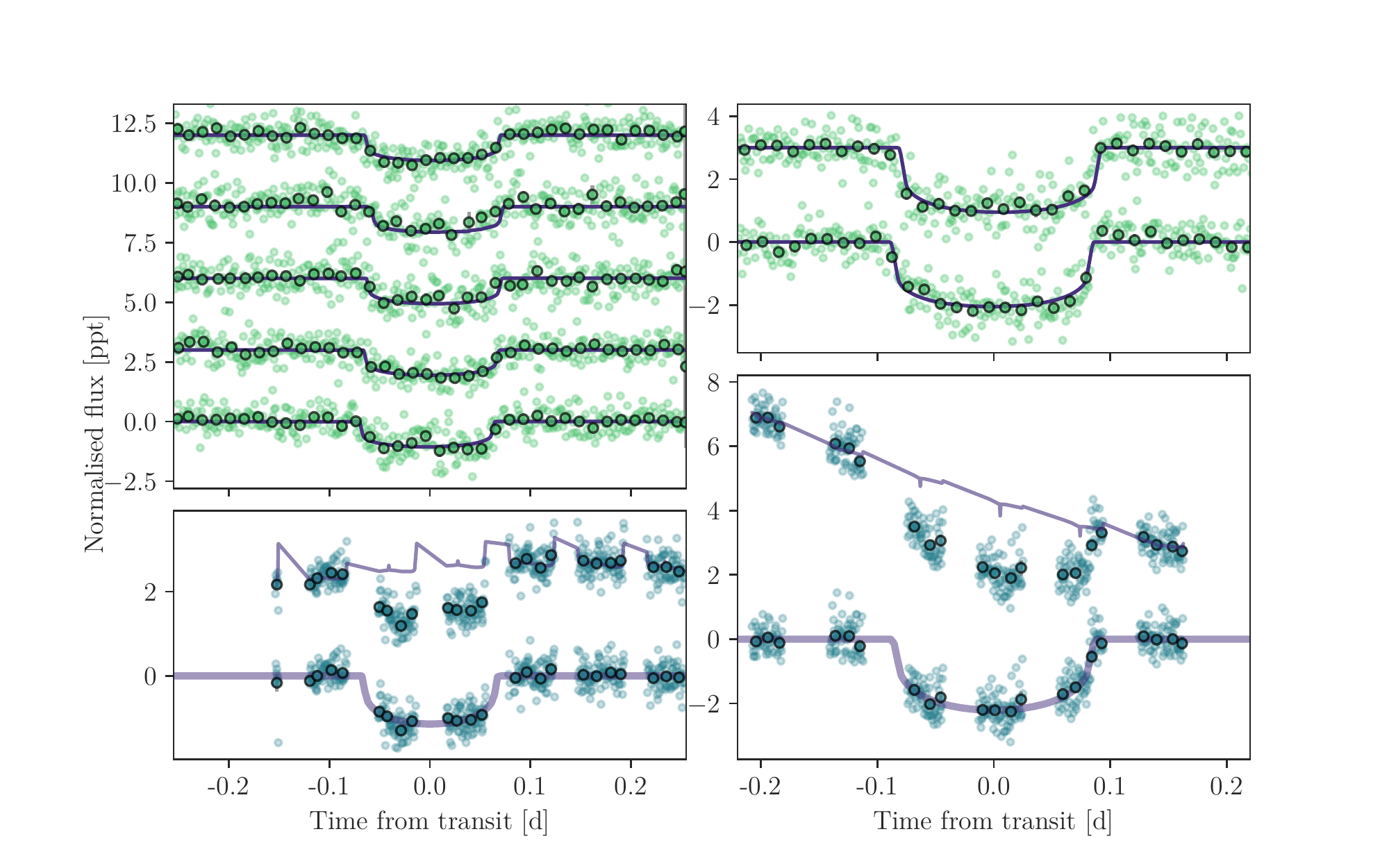}
      \caption{\tess{} (upper panels) and \cheops{} (lower panels) individual transits of planets b \& c. In the two lower panels, we show both the extracted \cheops{} flux with the best-fit decorrelation model (offset above), and the detrended \cheops{} flux with the best-fit transit model (below).
              }
         \label{fig:all_bc_trans}
   \end{figure*}

\cheops{} data unambiguously detected a unique $P=$\Tttvpercshort{}d period for TOI-2076~c (see the lower right panel of Figure \ref{fig:all_bc_trans}).
For TOI-2076~d, \refcomb{we have observed all aliases shorter than P=87.8\,d using either \cheops{} or ground-based facilities.}
The \cheops{} observations on 2021-04-29 and 2021-05-13 clearly ruled out the 43.9\,d alias, and 29.27\,d and 58.54\,d aliases, respectively (see second and fourth panels of Figure \ref{fig:35dflatlines}). 
\refcomb{Ground-based observations from LCO/McDonald covered the 25.1\,d alias, while photometry from both Saint-Ex and LCO/MuSCAT-3 covered the 35.1\,d alias.}

\refcomb{As our TESS-only models showed, the probabilities of periods longer than 80d\,d (87.8\,d and 175.6\,d)} are extremely low compared to close-in orbits due to both the period priors, and to the eccentricity priors derived from the transit shape.
The geometric and temporal period prior alone gives an 87.8\,d orbit a probability 28 times lower than that at 25.1\,d, while that at 175.6\,d is 179 times lower. For comparison the 35.1\,d orbit is disfavoured by only a factor 2.5.

We can therefore probabilistically exclude these \refcomb{longer orbits as well as those ruled out by \cheops{} observations} and focus only on the two short-period aliases for which we have ground-based observations - those at 25.1 and 35.1\,d. 

Our final combined model therefore has two goals -- provide accurate planetary parameters for all three planets, and determine the true period of the available planets. 
To do this, we modelled all available photometry simultaneously, including transit models for all three planets and detrending parameters.
For the competing period aliases, we built two models with identical parameters and changed only the period of TOI-2076~d. 
The relative difference in log likelihood can then be used for model selection, as the models otherwise share the same number of parameters \& datapoints.

\refcom{The size of the model means co-fitting the TESS light curve with a GP was not possible, therefore in order to remove residual systematic noise and/or stellar activity from the TESS light curve we subtracted a spline function fitted to the out-of-transit data and extrapolated over the transits.
We also masked outliers with flux $4\sigma$ away from both preceding \& succeeding points, and masked all points more than 3.5 transit durations from all transits to improve computational speed.}

The combined model was built using \texttt{PyMC3} \citep{exoplanet:pymc3}, which performs Hamiltonian Monte Carlo sampling - a far more efficient sampling technique than Markov Chain, as it can use the local gradient of the likelihood function to quickly move to distant regions of parameters space, even if correlated. 
Transit models used the \texttt{exoplanet} package \citep{2021JOSS....6.3285F}.

As our model contained 108 independent parameters, many of which included correlations, we were only able to sample the model thanks to first using the sampler provided with \texttt{exoplanet} (which is able to learn and explore off-diagonal covariances), and second by providing independent parameter groups on which to compute these covariances (namely, groups of detrending parameters for each telescope)\footnote{\url{https://dfm.io/posts/pymc3-mass-matrix/}.}. We ran each model with eight 3500-sample chains after a burn-in of 12\,000 steps to produce 28\,000 samples.
\refcom{We verified the gelmin-rubin statistic (\^{R}) was below 1.05, that the effective sample size was a large fraction of the total steps ($>10000$), and that the traces of individual chains were suitably mixed and Gaussian.}

\subsubsection{Treatment of \cheops{} data}
\cheops{} photometry can retain trends due to systematics, and previous works present in detail the techniques used to correct for these \citep[e.g.][]{bonfanti21,2021NatAs...5..775D,maxted2021analysis}.
We chose to co-fit the photometric transit models with a decorrelation against certain parameters. 
These included the x and y components of the roll angle $\sin{\Phi}$ and $\cos{\Phi}$, as well as estimates of background, CCD smear flux, and the change in temperature ($\Delta T$).
We also included both a linear and quadratic decorrelation component against time in order to model the stellar variability apparent in the \cheops{} lightcurves. 
\refcom{Past Cheops results \citep[e.g.][]{2021NatAs...5..775D,maxted2021analysis} have opted to further detrend as a function of roll-angle using for example a spline or Gaussian process fit, however our inspection of the Cheops flux residuals as a function of roll-angle for each visit revealed no apparent additional variations that would require such additional (and computationally intensive) modelling}.

\subsubsection{Treatment of ground-based data}
The majority of our ground-based observations are both affected by airmass trends and a lack of comparison stars.
On average, TOI-2076 provides 15 times more photons that all comparison stars combined, therefore relative photometry is dominated by the shot noise of comparison stars.
Instead we decided to use the raw aperture photometry and decorrelate against parameters linked to likely systematics.

Flux measurements with values $4\sigma$ above or below both their neighbours were masked.
In the case of MuSCAT-3 data, three differential colour time-series were derived using the normalised fluxes for all stars (target and comparisons) between neighbouring filters (e.g. g/r, r/i, i/z). 
We fitted a spline with 15-minute knot spacing to each of the four normalised flux timeseries (weighted by inverse photometric uncertainty), allowing us to interpolate differential colour estimates between each of the bands despite the asynchronous spacing of the four MuSCAT-3 detectors.
Given this is a confirmed multi-planet system without any nearby stellar companions, we can make the assumption that the transit depth should be unchanged across all filters, and therefore colour is independent of the transit.
This does not completely hold for limb-darkening, however this is a secondary effect \refcom{with the maximum difference in transit shapes between lightcurves being 120ppm and the average being only 50ppm - an order of magnitude smaller than the transit depth.
As this effect is dependent on a transit occurring, it cannot itself introduce a transit shape, and can only bias the derived parameters. 
Given the noise inherent in our ground-based data compared to e.g. the TESS transits, model parameters can only be minimally biased by this effect on the data. 
Therefore including colour information (which directly constrains colour-related systematics) as a linear decorrelation parameter results in a net improvement in the quality of the MuSCAT data and the general fit.}

To find the important detrending parameters, we first included a wide array of parameters in a local model including airmass, time, x and y centroid, width, full width half-maximum (FWHM), total comparison flux, and g/r, r/i and i/z (for Muscat-3).
These were normalised such that their medians were at 0.0 and the 1-sigma region spanned -1 to 1.
We then iterated multiple models, removing detrending parameters that resulted in statistically insignificant gradients.
The priors and posteriors for these detrending parameters are shown in Tables \ref{table:appendix_P35_model} and \ref{table:appendix_P35_model2}.

\refcom{Ground-based photometry (raw \& detrended) will also be made available through CDS.}

\subsubsection{Treatment of Limb Darkening}
We used the quadratic limb darkening parameters for all six bandpasses available.
In each case, we used theoretical limb darkening parameters calculated by \citet{2021RNAAS...5...13C} for \cheops{}, \citet{2018yCat..36180020C} for \tess{}, and \citet{2011A&A...529A..75C} for g, r, i, and z bandpasses.
In each case, we fitted a 2D interpolation surface to both $u_1$ and $u_2$ parameters as a function of \teff{} and \logg{}.
Then, using samples of TOI-2076 stellar parameters, the resulting distribution of limb darkening parameters were used to form a normal prior input to the transit models, although we rounded the prior standard deviation to 0.05 to avoid over-fitting.
   
  \begin{figure}
   \centering
   \includegraphics[trim={0 1.35cm 0 0.2cm}, width=\columnwidth]{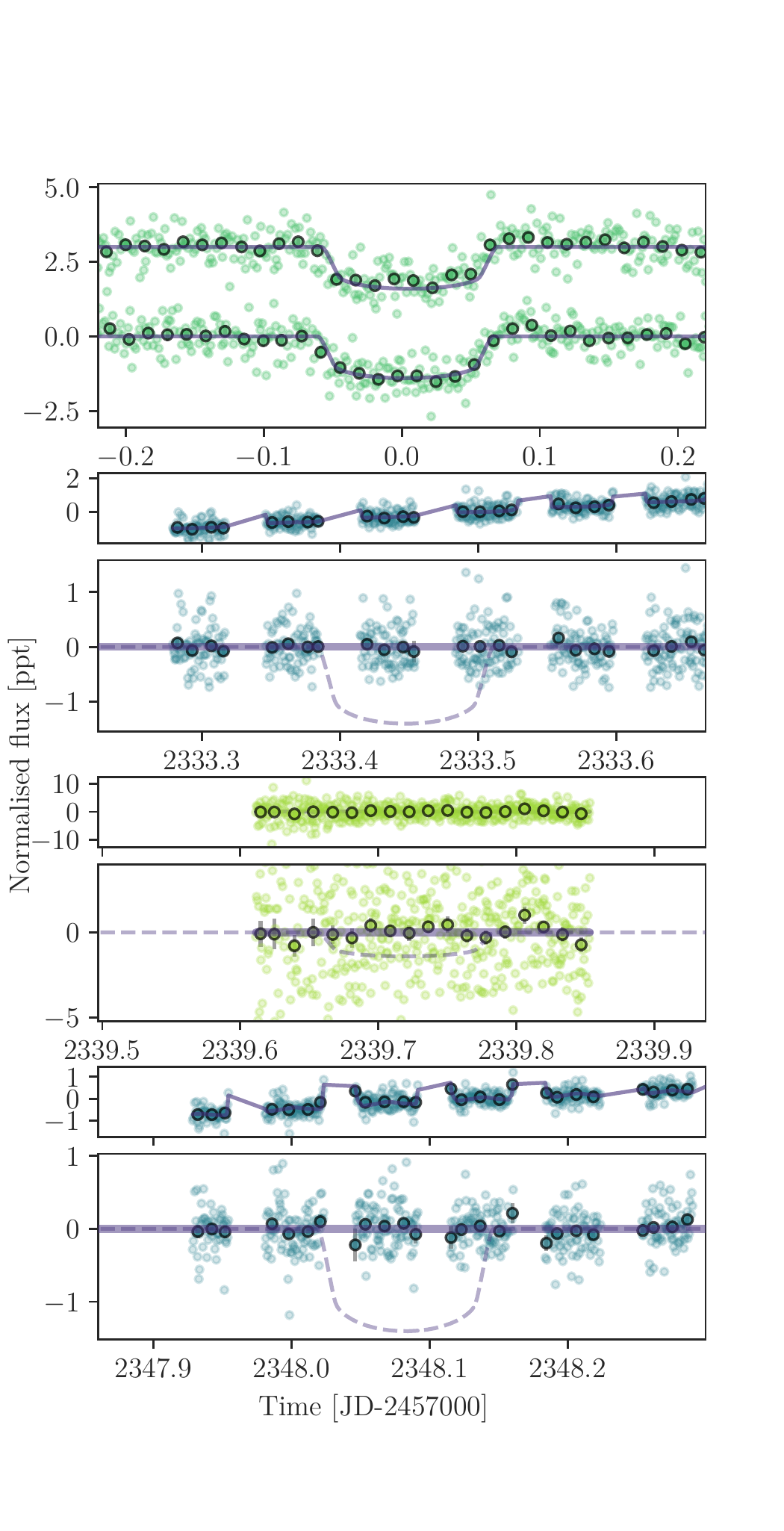}
      \caption{Observations of TOI-2076~d, including unsuccessful transit observations a) The two \tess{} transits detected by H21 as well as our best-fit model from the combined model. b) A \cheops{} observation covering the 43.9d alias. c) A LCO/McDonald 1m Sinistro lightcurve of the 25.09d alias. d) A \cheops{} observation covering both 29.27d and 58.54d aliases. In the lower three panels, a TOI-2076~d model-fit is shown to demonstrate the expected transit shape \& depth. In the lower three plots, the upper points \& line show the raw flux \& best-fit decorrelation model, while the lower panel shows the detrended flux \& the expected transit model.
              }
         \label{fig:35dflatlines}
   \end{figure}
   
  \begin{figure}
   \centering
   \includegraphics[trim={0 1.35cm 0 0.2cm}, width=\columnwidth]{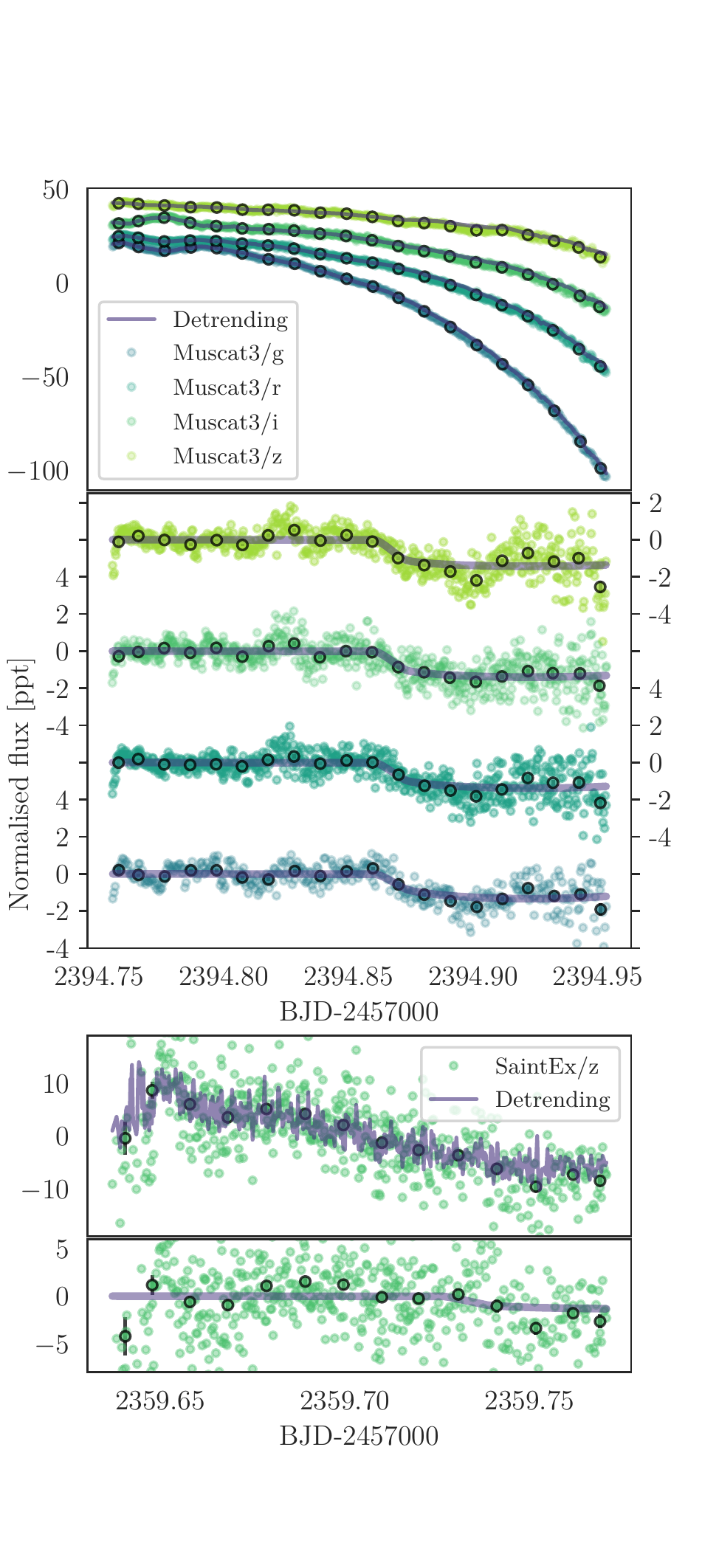}
      \caption{Ground-based observations of the 35.1d alias of TOI-2076~d. a) Raw LCO/MuSCAT-3 observations of TOI-2076 in four filters. b) Detrended lightcurves in the four filters, along with best-fit transit models. c) Raw SAINT-EX photometry in $r$ band. d) Detrended SAINT-EX photometry along with a best-fit transit model.
              }
         \label{fig:ground_transit_35d}
   \end{figure}

\subsection{TTV analysis}
\label{sect:TTVs}
\begin{figure}
   \centering
   \includegraphics[width=\hsize]{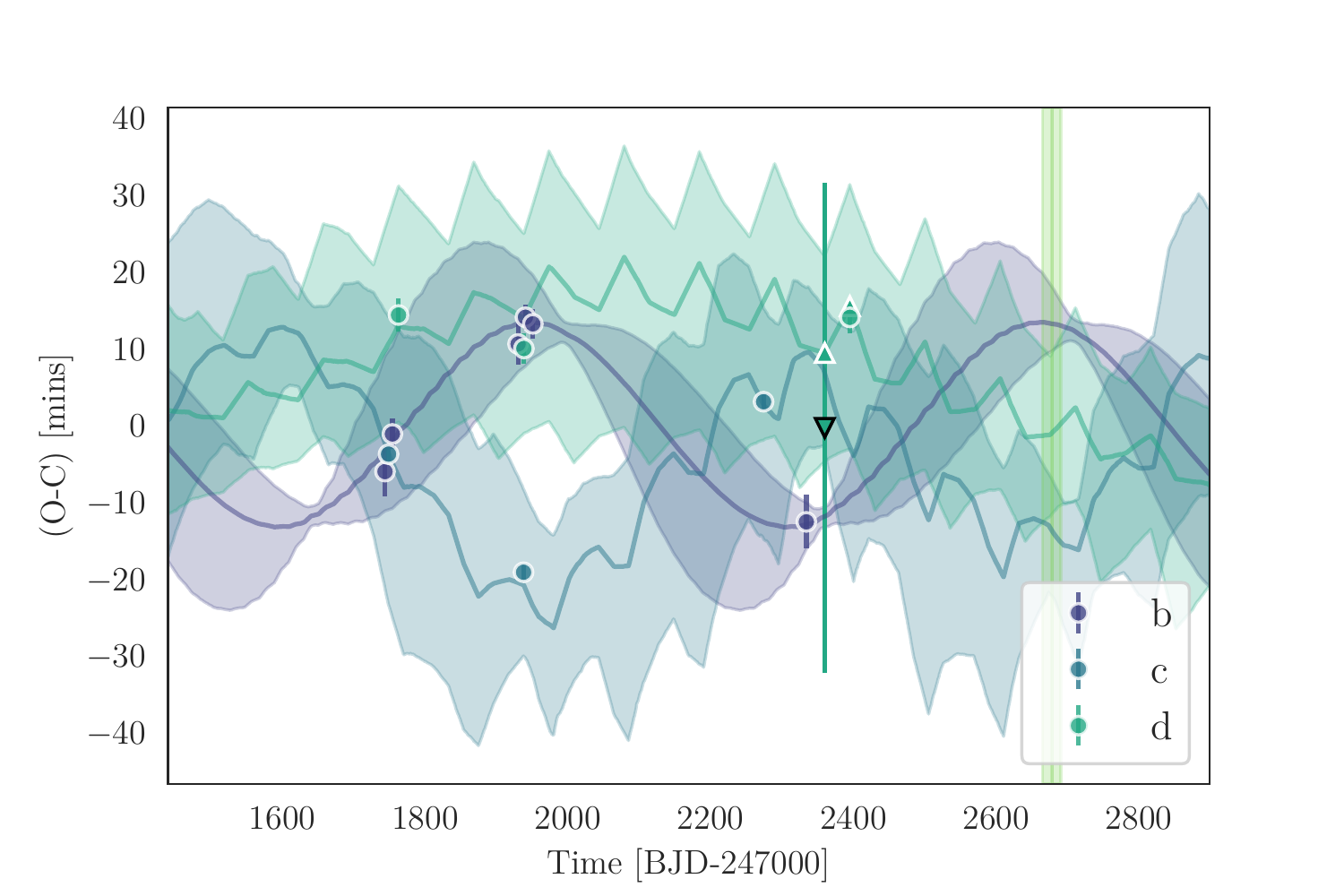}
      \caption{Observed TTVs and \texttt{TTVFaster}/\texttt{Ultranest} TTV models for each of the three planets. Coloured lines show the median best-fit TTV models, while coloured regions show 1-$\sigma$ range. The predicted transit time for the low S/N SAINT-EX observation of planet d is shown as a triangle with white edges, while the observed transit time is shown with dark-edged inverted triangles. Planned TESS observations of the two planets in 2022 are shown in light green.}
         \label{fig:TTVmodels}
   \end{figure}

Rather than fitting for a specific fixed period, our combined model fitted transits individually using a normal prior centred on the expected time of transit given a linear ephemeris \refcom{and a loose standard deviation of 0.025d (36 mins)}.
These outputs revealed clear TTVs, with the \cheops{} transit of TOI-2076~b arriving $57\pm5$ minutes early compared to a linear ephemeris using only the \tess{} data, while TOI-2076~c arrived $50\pm4$ minutes late.
\refcom{This can be seen in Figure \ref{fig:TTVmodels}.}
TTVs are also expected given the period ratios of the planets are close to period commensurability.

To analyse the observed TTVs and ensure confidence in our results, we performed independent TTV analyses using three distinctly different approaches.
The same derived transit times and errors were then used as input to these analyses.
As the SAINT-EX ground-based lightcurve of TOI-2076~d has a low S/N, it is excluded from the TTV analysis.

\refcom{We performed three approaches - one using the \texttt{TTVfaster} package \citep{2016ascl.soft04012A} and \texttt{Ultranest} Nested Sampling \citep{2021JOSS....6.3001B}, and two using the approach presented in \cite{2021A&A...655A..66L} which used the {\ttfamily TTVfast} algorithm \citep{ttvfast2014}, and the \texttt{samsam}\footnote{\url{https://gitlab.unige.ch/Jean-Baptiste.Delisle/samsam}.} MCMC algorithm \citep[see][]{Delisle2018}.
For full details of these fits, see Section \ref{app:ttv}.
The output of the first model (\texttt{TTVfaster}/\texttt{Ultranest}) is shown purely for reference in Figure \ref{fig:TTVmodels}.}

%   \begin{figure}
%   \centering
%   \includegraphics[width=\hsize]{4dtrans_ultranest_analysis_goodinputs_nojitter2mass_kdes_wpriors.pdf}
%       \caption{KDEs of planetary mass posteriors for the three planets in TOI-2076 as derived from the \texttt{TTVFaster}/\texttt{Ultranest} model fits. Dashed lines show the broad mass priors for each, derived from planetary radii.}
%          \label{fig:TTVmasses}
%   \end{figure}
%@arxiver{all_bc_transits_ref.pdf,ground_35d_data2.pdf,Pre-Cheops_modelling_bc_postref.pdf}

\subsection{Orbital stability analysis}

In order to test the compatibility of the two remaining high-probability aliases at 25.1\,d \& 35.1\,d with the 21\,d period of TOI-2076~c, we performed an N-body stability analysis. We used the \texttt{rebound}\footnote{\url{https://github.com/hannorein/rebound}.} package
\citep{Rein2012A&A...537A.128R} 
with the \texttt{whfast} integrator \citep{Rein2015MNRAS.452..376R} and
we activated the Mean Exponential Growth factor of Nearby Orbits \citep[MEGNO,][]{Cincotta2000A&AS..147..205C} indicator.
The orbital configuration can be considered stable when the value of MEGNO is close to 2
\citep{Cincotta2000A&AS..147..205C}.
We compute the masses using \texttt{forecaster}\footnote{\url{https://github.com/bmorris3/forecaster} python3 version.}
from the planetary radii and stellar mass and radius.
We used the radii values of planet b and c determined in Section~\ref{sect:premod} and \ref{sect:combmod},
we assumed an error of $0.5\, R_\oplus$ on radius of the planet d.
We drew 1000 values of masses between the lower and upper boundaries provided by \texttt{forecaster}, 
and assumed uniform distribution of the mean anomaly (between 0 and $2\pi$), for each planet.
We fixed the eccentricity to 0, argument of pericenter to $90^{\circ}$,
and the longitude of the ascending node to $180^{\circ}$, for each planet.
We sampled 500 values of the period of the planet d for each of the two possible aliases,
from two Gaussian distribution centred at 25.09\,d and 35.126\,d, both with a standard deviation of 0.1\,d \refcom{- chosen to be larger than both the period uncertainties ($<0.0001$d) and the observed TTVs (0.02\,d) to guard against systematic uncertainties.}
The inclinations and periods have been assumed normally distributed with the values and uncertainties obtained from the  analysis described in Section~\ref{sect:premod} and \ref{sect:combmod}.
We ran 1000 simulations and we integrated the orbits for $100\ 000$ years with a step-size of 0.25\,d. 
\refcom{We assigned a value of the MEGNO indicator $\gg 2$ if the system underwent a close encounter or if a body gained a semi-major axis greater than 150~au.}

\section{Results}
\subsection{Combined Model}
\refcomb{The derived planetary parameters from our combined model can be seen in Table \ref{table:planet_params}.}
We find planetary radii of \Trplzero{}, \Trplone{}, and \Trpltwo{}\,$R_\oplus$, respectively, which are significantly smaller than those of H21 which found $3.282\pm0.043$, $4.438 \pm 0.046$, \& $4.14 \pm0.07 R_\oplus$.
The main reason for this appears to be due to a bug in the modelling performed in H21 where the radius ratio ($R_p/R_s$) was submitted to \texttt{exoplanet}'s \texttt{LimbDarkLightCurve} function, rather than the radii in solar units ($R_p/R_\odot$). 
This led to final radius values that were inflated by a factor of $R_\odot/R_s \sim1.31$.
Hence, the radii and radius ratios defined here should supersede those in H21.

\begin{table*}
\centering                          % used for centering table
\caption{Derived planetary parameter posterior distributions for each of the three planets.}             % title of Table
\label{table:planet_params}      % is used to refer this table in the text
\begin{tabular}{l c c c}        % centred columns (4 columns)
\hline\hline                 % inserts double horizontal lines
Parameter & TOI-2076~b & TOI-2076~c & TOI-2076~d \\
\hline                        % inserts single horizontal line
%Linear period, $P$ [d] & \Tderivedperiodzero & \Tderivedperiodone & \Tderivedperiodtwo \\
Epoch, $t_0$ [BJD-2457000] & \Ttransittimesonezero & \Ttransittimesoneone & \Ttransittimesonetwo \\
Period, $P$ [d] & \Tttvperb & \Tttvperc & \Tttvperd \\
Semi-major axis, $a$ [AU] & \Tsmazero & \Tsmaone & \Tsmatwo \\
Radius ratio, $R_p/R_s$ & \Trorzero & \Trorone & \Trortwo \\
Duration, $t_{D}$ [hrs] & \Ttdurzero & \Ttdurone & \Ttdurtwo \\
Radius, $R_p$ [$R_\oplus$] & \Trplzero & \Trplone & \Trpltwo \\
Insolation, $I_p$ [$Wm^{-2}$] & \TSinzero & \TSinone & \TSintwo \\
Surface Temp., $T_{\rm eq}$ [K] & \TTsurfpzero & \TTsurfpone & \TTsurfptwo \\
%TTV Masses, $M_p/M_{\oplus}$ [K] & \TttvMb & \TttvMc & \TttvMd \\
%Equilibrium Temp., $T_{\rm eq}$ [K] &\TTeqpzero & \TTeqone & \TTeqptwo \\
TSM & \TttvTSMb$^{\star}$ & \TttvTSMc$^{\star}$ & \TttvTSMd$^{\star}$ \\
\hline                                   %inserts single line
\end{tabular}
\begin{tablenotes}
\small
\item $^{\star}$ - The TSM values are calculated using our tentative \texttt{TTVFaster}/\texttt{Ultranest} models which are typically $<1\sigma$ from the predictions of \citet{chen2016probabilistic}, but true masses will require more observations.
\end{tablenotes}
\end{table*}

\refcomb{As shown in Section \ref{sect:combmod}, \cheops{} photometry clearly reveals the True period of TOI-2076c to be \Tttvperc{}.
All period aliases shorter than 87.8\,d were observed either with \cheops{} or from the ground.
As shown in our TESS-only analysis (Sect \ref{sect:premod}), the longest period aliases (87.8\,d \& 175.6\,d) are orders of magnitude less likely due to constraints from the lightcurve as well as period \& eccentricity priors.
\cheops{} photometry clearly rules out the 29.27\,d, 43.9\,d and 58.54\,d aliases.
This left the 25.1\,d and 35.1\,d aliases for which we obtained ground-based photometric follow-up.}

\refcomb{In order to assess the fit and implications of our model fits to each period alias, we computed the differences in log likelihoods, log priors and log probabilities in Table~\ref{tab:logliks}. As the number of data points and parameters are preserved across models, the difference in Bayesian information criterion \citep[$\Delta$ BIC,][]{10.1214/aos/1176344136} is simply $\Delta{\rm BIC} = -2 \Delta \log{\rm prob}$.
Typically $\Delta{\rm BIC}>10$ or $\Delta\log{\rm prob}<-5$ is used to show strong support for a model \citep{raftery1995bayesian}.}

\refcomb{Our combined model clearly prefers the 35.1\,d alias as opposed to the 25 and 88\,d aliases, as can be seen from the derived log probability differences in Table~\ref{tab:logliks}.
When combined with the log-priors derived from the combination of geometric, window function and eccentricity priors derived in our TESS-only modelling, we find $\Delta \log{\rm prob}$ values of more than 150 in favour of the $P=35.1$\,d model.
We therefore adopt this as the true period of TOI-2076\,d, although we further discuss the orbit of TOI-2076 in \ref{sect:d_orbit}.}

\begin{table}
\centering                          % used for centering table
\caption{Log probabilities for each of the remaining aliases. Log likelihood from the combined model fit described in section \ref{sect:combmod}, log priors from the initial modelling described in section \ref{sect:premod}. \refcom{The final column shows the difference in $\log{\rm prob}$ with respect to the $P=35.1$d model (i.e. $\log{p_{{\rm per},i}} - \log{p_{35}}$). Here higher $\Delta\log{\rm prob}$ is associated with the best-fitting model.}}             % title of Table
\label{tab:logliks}      % is used to refer this table in the text
\begin{tabular}{l c c c}        % centred columns (4 columns)
\hline\hline                 % inserts double horizontal lines
Period & Log likelihood & log prior & $\Delta \log{\rm prob}$ \\
\hline                        % inserts single horizontal line
%Linear period, $P$ [d] & \Tderivedperiodzero & \Tderivedperiodone & \Tderivedperiodtwo \\
25.1d & -3003.2 & -215.4 & -291.4 \\
35.1d & -2908.8 & -18.4 & \textbf{0.0} \\
87.8d & -3074.9 & -20.7 & -168.3 \\
175.6d & -3074.9 & -53.6 & -201.3 \\
\hline                                   %inserts single line
\end{tabular}
\end{table}

\subsection{TTVs}

\refcom{We find for the first time that the TOI-2076 system exhibits large TTVs with amplitudes greater than 20 minutes for planets b \& c.
However, our three approaches to modelling the TTVs each find inconsistent planetary masses (see Table \ref{tab:ttvres} \& Section \ref{app:ttv}).}
This implies that the number of transit timing measurements is not yet sufficient to obtain robust mass estimates from TTVs and, as expected for a model without strong constraints from the data, the choice of prior modifies the resulting posteriors, as can be seen in the determined masses and eccentricities in Table \ref{tab:ttvres}.
We therefore caution use of those parameters derived from TTVs (i.e. planetary mass) until more transits can be observed,\refcomb{ although we use the \texttt{TTVFaster/Ultranest} results (which have the most realistic prior distributions and output masses) as representative masses for future calculations (e.g. TSM)}.

\refcom{The best-fitting models do appear to suggest a significant anti-correlated TTV signal} between planets b and c, due to their proximity to the 2:1 mean motion resonance creating a \Tttvbcsuperper{}d super-period.
The relationship between planets c and d is not well defined due to the number of transits, \refcom{but our best-fit TTV models suggest that long-term sinusoidal TTVs between planets c \& d could} be observed in the future, as well as a potential chopping signal that could allow for precise mass measurements.

\subsection{Orbital stability}
We found the $5.3\pm1\%$ of the simulations around the 25.1-d-alias are stable (MEGNO $~\sim 2$),
while the $89\pm1.4\%$ of the simulations (445 out of 500) around the 35-d-alias are stable.
These results indicate that the 35-d-alias is the most favourable period for the planet d.

%--------------------------------------------------------------------
\section{Discussion}

\subsection{The orbit of TOI-2076~d}\label{sect:d_orbit}
\refcomb{In Table \ref{tab:logliks} we revealed the differences in log probabilities between three difference period aliases.
The major difference in log likelihood are driven by the presence and absence of transits in ground-based follow-up data. 
The TESS data, which only allowed for identifying the original period aliases, consequently show near-identical transit models and log-likelihoods.}
For the 25d case \refcomb{the loglikelihoods are the result} of a transit in the LCO/McD data, but a flat line in the MuSCAT3 data; the 35d case is the loglikelihood of a flat line in the LCO/McD data and a transit in MuSCAT3; while the 88 \& 176d aliases show the loglikelihood of flat lines in each of the ground-based transits.

\refcom{The largest difference in log likelihood ($\sim100$) comes from the LCO/MuSCAT-3 observations.}
Despite the fact that the LCO/MuSCAT-3 data required substantial detrending with respect to colour and airmass, the transit model was far better able to explain the sharp ingress feature at BJD=2459395.87 compared to linear detrending, which occurred precisely at the expected transit time given a linear ephemeris (upper panels, Figure \ref{fig:ground_transit_35d}). 
%and the detrending parameters were flexible enough to absorb some of the flux drop in the $P_d=25$ model, 
%Indeed, although the $P_d=25.1$d was able to detrend the in-transit points to the median flux (likely due to parameters correlating well with time), there still appears to be a clear ingress feature in that dataset which is not detrended (see Figure \ref{fig:P25P35difference}).
The SAINT-EX data, which was lower S/N and covers only a very short duration of in-transit data, is not as conclusive as the LCO/MuSCAT-3 data, although the transit model is also marginally preferred in this dataset (lower panel, Figure \ref{fig:ground_transit_35d}).

The second reason for the better model (\refcom{a difference of $\sim75$ in log likelihood}) fit is the non-detection of a transit using the LCO/1m data from McDonald observatory during a purported transit of the $P_d=25$\,d alias. 
The observation, which occurred at low airmass and covered the entire expected transit event, appears to see no clear flux drop, and a flat model is preferred over a transit one (lower panel, Figure \ref{fig:35dflatlines}).

This hypothesis is also supported by our orbital stability analysis - where the vast majority of long-term orbits for the $P_d=35.1$\,d alias are stable, and those of the $P_d=25.1$\,d orbit are not.
However, the $P_d=25.1$\,d scenario could be stable if planets c and d were caught in an MMR, for example the 6:5 configuration which, although less common than the 5:3 ratio implied by the $P_d=35.1$\,d, is not impossible \citep[e.g. the Kepler-36 system][]{2012Sci...337..556C}.
Such a possibility seems less likely given the potentially disturbing influence of the inner $P=$\Tttvperb{} planet (which, as discussed in Sect.~\ref{sect:TTVs}, is not in resonance), and given the fact that the observed TTV of planet c appears satisfactorily explained by anti-correlation with the TTVs of planet b, rather than due to the influence of any closer-proximity outer planet.

The 35.1\,d alias also appears more likely when considering the orbital periods of the system.
Planets b \& c are close to but slightly outside of a 2:1 period ratio (2.03).
Such a pile-up of planets just beyond period ratios is common in multi-planet systems and may be a hallmark of disc migration \citep{2014ApJ...790..146F}.
The 35.1d alias follows that trend by being extremely close but just outside of a 5:3 orbital ratio with planet c (5.014:3).
None of the other potential period aliases show this pattern, although the $P=25.08936$\,d alias is just inside a 6:5 ratio (5.969:5).

Taken together, we believe the evidence for a 35.1\,d period for TOI-2076~d is compelling and we hereafter refer to it as the correct period.

\subsection{Planetary Characteristics}
%Although our TTV models show degeneracies, they do hint that the three planets transiting TOI-2076 are at the lower end of typical masses for the observed planetary radii.
%From the radii alone, we can say that all three likely have extended H-He atmospheres.
%, with planets c \& d likely potentially hosting as much as $\sim 3$\% by mass.

%   \begin{figure}
%   \centering
%   \includegraphics[width=\hsize]{MR_young.pdf}
%       \caption{Mass-radius distribution for sub-Neptunes with TOI-2076. Grey points with errors show planets with well-defined mass measurements from the literature. Shaded regions shows a simple kernel density estimate for the most-populated regions of the parameter space. Straight lines show solid-body internal structure models with earthlike cores and 0, 50, \& 100\% H$_2$O respectively. Curved lines show earthlike cores with 0.1, 1, \& 2\% H$_2$ (for an equilibrium temperature of 700K). All internal composition models are from \citet{2019PNAS..116.9723Z} \footnote{\url{https://lweb.cfa.harvard.edu/~lzeng/planetmodels.html}}.}
%          \label{fig:MR}
%   \end{figure}
   
Thanks to our determination of planetary periods, we now know that the TOI-2076 planets are irradiated by 84, 32, and 16\,$S_\oplus$ respectively.
Compared to many sub-Neptunes so-far detected, this is remarkably low and suggests that the effect of stellar insolation on e.g. their radii must be minimal.
From the radii alone, we can say that all of the three TOI-2076 planets likely have extended H-He atmospheres.
These inflated radii may in part be explained by their youth.
Young planets are affected by a handful of processes which could change their bulk physical parameters.
The first is photo-evaporation, however with a star of age $340\pm80$\,Myr and orbits of $>0.05$\,AU, this is likely no longer a dominant effect except potentially for planet b.
Also important is the process of core-powered mass-loss by which small planets with light gaseous envelopes can lose their outer layers through thermal heating by the cooling core \citep{2018MNRAS.476..759G}. Finally, atmospheric contraction may still be acting on the TOI-2076 planets \citep[e.g.][]{2012ApJ...761...59L}.
\citet{2020AJ....160..108B} explored differences in radius populations as a function of planetary age, and found that the average radius of sub-Neptunes appears to shift with time from $\sim$3.0\,$R_\oplus$ at $<1$\,Gyr to $\sim$2.5\,$R_\oplus$ at $>1$\,Gyr, particularly for planets with irradiation less than 150$\,S_\oplus$ like TOI-2076~b, c, and d.
With radii of $R_c=$\Trplone{} and $R_d=$\Trpltwo{}$\,R_\oplus$, the outer planets in the TOI-2076 system may provide evidence that young sub-Neptunes are born with even more inflated radii than the $\sim$3.0\,$R_\oplus$ seen in \citet{2020AJ....160..108B}.
If puffy H-He envelopes are able to be maintained for hundreds of Myr, it could be a sign that core-powered mass loss and/or contraction are slower processes than previously thought. 
The atmospheres of the outer planets orbiting TOI-2076~could therefore be the perfect test-beds for such theories.
   
\subsection{Future observations}

TOI-2076 will be re-observed by \tess{} in Sector 50 (2022-Mar-26 to 2022-Apr-22; see Figure \ref{fig:TTVmodels}).
Although exact downlink gaps are not yet known, it is likely that b will show 3 transits, while both c and d will transit once.
The timing of these transits will help further constrain TTVs for this system, and can be helped by a campaign of observations with \cheops{}, especially to observe sequential transits of c and d, thereby potentially detecting the predicted chopping signal and better constraining the masses of the three planets.

We predict expected RV semi-amplitudes of \TttvKb{}, \TttvKc{}, \& \TttvKd{}\ms{} \refcom{using the \refcomb{provisional} masses implied by our \texttt{TTVFaster}/\texttt{Ultranest} models (although we caution that robust TTV masses will require more transit observations}).
This would make these three planets extremely challenging targets, especially when considering the strong $\sim$7\,d rotation signal present in the \tess{} light curve. Therefore, TTVs may prove the best method of constraining the planetary masses for the three planets around TOI-2076.
  \begin{figure}
  \centering
  \includegraphics[width=\hsize]{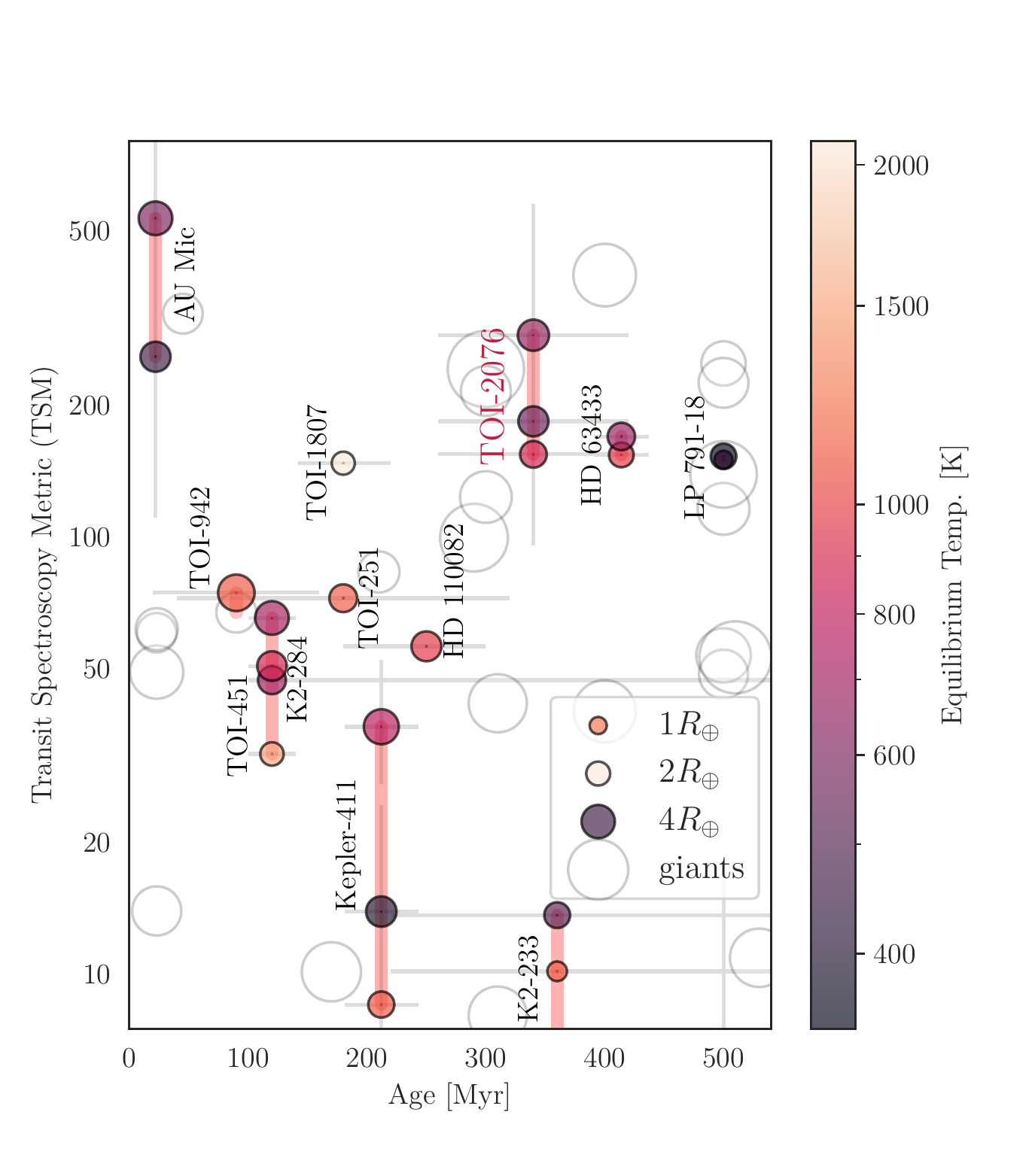}
      \caption{Comparison of transmission spectroscopy metrics for all small transiting planets ($R_p<5R_\oplus$) around young stars (age$<$500\,Myr) as a function of age in million years (Myr). Point size represents planetary radius, while point colour shows equilibrium temperature. Multi-planet systems with small planets are connected by red trails. The TOI-2076 system is labelled in red \refcom{with TSM calculated using \refcomb{tentative} masses derived from our \texttt{TTVFaster}/\texttt{Ultranest} models.}}
         \label{fig:TSMs}
  \end{figure}
Of the many small young planets detected by \tess{}, TOI-2076 hosts three of the most atmospherically accessible, all with transmission spectroscopy metrics \citep[TSM;][]{2018PASP..130k4401K} above 100, although those values host large uncertainties due primarily to the large mass uncertainties.
Indeed, if the mass of TOI-2076~c is confirmed to be \TttvMcshort{}$M_\oplus$, it is amongst the highest-ranked cool \& small planets with $T_{\mathrm{eq}}<750$\,K, $R_p<4\,R_\oplus$ found by TESS.
There are also very few small young planets with equivalently accessible atmospheres (see Figure \ref{fig:TSMs}), with only the mini-Neptunes around AU Mic \citep{2020Natur.582..497P} and HD\,63433 \citep[TOI-1726,][]{2020AJ....160..179M} with similar TSM values, likely due to their host stars superior brightnesses (G = 7.8 \& 6.7).
Hence, the planets around TOI-2076 could form key targets for future atmospheric follow-up with e.g. the James Webb Space Telescope (JWST).
Their long periods may also mean the outer planets are relatively unperturbed by stellar radiation pressure or wind, enabling the planets to maintain large exospheres which may be detectable in the UV with e.g. the STIS or COS instruments on the Hubble Space Telescope.
%AU Mic = 50, TOI-1726 (mini-Neptunes)=193, 212, TOI-451: 43,47,58, HD110086 (3.3Re): 67

%--------------------------------------------------------------------
\section{Conclusions}
We performed targeted follow-up photometry of the period aliases of the young, long-period sub-Neptunes TOI-2076~c and d in order to confirm their orbital periods and further characterise the system.
We initially modelled the planetary system using \texttt{MonoTools}, developed specifically for this task, which is able to take available stellar and photometric data and calculate a marginal probability for each period alias.

Using ESA's \cheops{} space telescope, we performed targeted follow-up of the highest-probability aliases and were able to confirm the \Tttvperc{}\,d period alias as the true one for TOI-2076~c.
\cheops{} observations also helped rule out three of the most-probable period aliases for TOI-2076~d.
Ground-based photometry from the 1-meter LCO/Sinistro telescope at McDonald Observatory enables us to discard one of the remaining high-probability aliases for TOI-2076~d at 25.1\,d, which is also hinted at by stability \& TTV analyses of this alias in the presence of the \Tttvpercshort{}\,d TOI-2076~c.
Furthermore, ground-based observations with both the 2-meter LCO/MuSCAT-3 on the Faulkes North telescope at Haleakala and the 1-meter SAINT-EX telescope at San Pedro Martir were able to detect ingresses of the 35.1\,d alias. Bayesian model comparison vastly favoured this alias over the other unexcluded aliases ($\Delta\log{p}>100$), confirming \Tttvperd{}d as the true period of TOI-2076~d.

With high-precision space-based transit observations spanning two years, thanks to \tess{} and \cheops{}, we were able to improve the ephemerides and radius precision, with updated radii of $R_b=$\Trplzero{}, $R_c=$\Trplone{}, and $R_d=$\Trpltwo{}$R_\oplus$.
These transits also enabled us to detect anti-correlated TTVs between TOI-2076~b \& c with an amplitude of $\sim$30\,min\refcom{, although TTV modelling did not have enough observed transits to constrain masses \& eccentricities.
The three planets inflated radii} suggest all three are low-density warm sub-Neptunes with significant hydrogen envelopes, potentially still undergoing atmospheric contraction.
Their large radii, low incident flux and bright host star magnitude make all three planets extremely interesting targets for future atmospheric characterisation with e.g.\ JWST.

\begin{acknowledgements}
\refcomb{We thank the anonymous referee for a thorough and constructive review.}
\cheops{} is an ESA mission in partnership with Switzerland with important contributions to the payload and the ground segment from Austria, Belgium, France, Germany, Hungary, Italy, Portugal, Spain, Sweden, and the United Kingdom. The \cheops{} Consortium would like to gratefully acknowledge the support received by all the agencies, offices, universities, and industries involved. Their flexibility and willingness to explore new approaches were essential to the success of this mission.
This work has been carried out within the framework of the NCCR PlanetS supported by the Swiss National Science Foundation.
This material is based upon work supported by the National Science Foundation through the US Community Open Access to LCO facilities provided under Grant No. 2034337.
This paper is based on observations made with the MuSCAT3 instrument, developed by Astrobiology Center and under financial supports by JSPS KAKENHI (JP18H05439) and JST PRESTO (JPMJPR1775), at Faulkes Telescope North on Maui, HI, operated by the Las Cumbres Observatory.
This research made use of \textsf{exoplanet} \citep{exoplanet:joss,
exoplanet:zenodo} and its dependencies \citep{exoplanet:agol20,
exoplanet:arviz, exoplanet:astropy13, exoplanet:astropy18, 2013MNRAS.434L..51K,
exoplanet:luger18, exoplanet:pymc3, exoplanet:theano}.

%%%%%%%%%%%%%%%%%%%%%%%%%%%%%%%%%%%%%%%%%%%%%%%%%%%%%%%
%.    WARNING:  DO NOT EDIT ACKNOWLEDGEMENTS BELOW    %
%%%%%%%%%%%%%%%%%%%%%%%%%%%%%%%%%%%%%%%%%%%%%%%%%%%%%%%
%
% Author info & acknowledgements are auto-generated from this google spreadsheet: https://docs.google.com/spreadsheets/d/17KRQMYalFKKTm0ZJO5wHxUhVfgf1chrhw8q7WHVYig0/edit?usp=sharing 
%
This work makes use of observations from the LCOGT network. Part of the LCOGT telescope time was granted by NOIRLab through the Mid-Scale Innovations Program (MSIP). MSIP is funded by NSF.\\ %K. A. Collins
AT acknowledges support from an STFC PhD studentship.\\ %A. Tuson
This work has been carried out within the framework of the NCCR PlanetS supported by the Swiss National Science Foundation.\\ %S. Ulmer-Moll
DG and LMS gratefully acknowledge financial support from the \emph{Cassa di Risparmio di Torino} (CRT) foundation under Grant No. 2018.2323 ``Gaseousor rocky? Unveiling the nature of small worlds''.\\ %D. Gandolfi
S.G.S acknowledges the support from FCT through Investigador FCT contract nr. CEECIND/00826/2018 and  POPH/FSE (EC). This  work  was  supported  by FCT - Fundac\c{c}\~ao  para  a  Ci\^encia  e  Tecnologia through national  funds and by FEDER through COMPETE2020 - Programa Operacional Competitividade e Internacionaliza\c{c}\~ao by these grants:PTDC/FIS-AST/32113/2017 \& POCI-01-0145-FEDER-032113; PTDC/FIS-AST/28953/2017 \& POCI-01-0145-FEDER-028953; PTDC/FIS-AST/28987/2017 \& POCI-01-0145-FEDER-028987.\\ %S.G. Sousa
ACC and TGW acknowledge support from STFC consolidated grant numbers ST/R000824/1 and ST/V000861/1, and UKSA grant ST/R003203/1. \\ %T.G. Wilson
V.V.G. is a F.R.S.-FNRS Research Associate. The Belgian participation to CHEOPS has been supported by the Belgian Federal Science Policy Office (BELSPO) in the framework of the PRODEX Program, and by the University of Li\`ege through an ARC grant for Concerted Research Actions financed by the Wallonia-Brussels Federation.\\ %Valerie Van Grootel
LBo, GPi, VNa, GSs, IPa, and RRa acknowledge the funding support from Italian Space Agency (ASI) regulated by ``Accordo ASI-INAF n. 2013-016-R.0 del 9 luglio 2013 e integrazione del 9 luglio 2015 CHEOPS Fasi A/B/C''.\\ %L. Borsato
MJH and YA acknowledge the support of the Swiss National Fund under grant 200020\_172746.\\ %Matthew J. Hooton
M.L. acknowledges support of the Swiss National Science Foundation under grant number PCEFP2\_194576.\\ %M. Lendl
ABr was supported by the SNSA.\\ %A. Brandeker
B.-O.D. acknowledges support from the Swiss National Science Foundation (PP00P2-190080).\\ %Brice-Olivier Demory
PM acknowledges support from STFC research grant number ST/M001040/1.\\ %Pierre F. L. Maxted
This work is partly financed by the Spanish Ministry of Economics and Competitiveness through grants PGC2018-098153-B-C31\\ %Enric Pallé
Acknowledges support from the Spanish Ministry of Science and Innovation and the European Regional Development Fund through grant PGC2018-098153-B- C33, as well as the support of the Generalitat de Catalunya/CERCA programme.\\ %Ignasi Ribas
This project was supported by the CNES\\ %Magali Deleuil
M.G. is an F.R.S.-FNRS Senior Research Associate.\\ %Michaël Gillon
GyMSz acknowledges the support of the Hungarian National Research, Development and Innovation Office (NKFIH) grant K-125015, a a PRODEX Experiment Agreement No. 4000137122, the Lend\"ulet LP2018-7/2021 grant of the Hungarian Academy of Science and the support of the city of Szombathely.\\ %Gyula M. Szabó\\ %Gyula M. Szabó
This work was granted access to the HPC resources of MesoPSL financed by the Region Ile de France and the project Equip@Meso (reference ANR-10-EQPX-29-01) of the programme Investissements d'Avenir supervised by the Agence Nationale pour la Recherche\\ %Jacques Laskar
L.D. is an F.R.S.-FNRS Postdoctoral Researcher.\\ %Laetitia Delrez
MF gratefully acknowledge the support of the Swedish National Space Agency (DNR 65/19, 174/18).\\ %Malcolm Fridlund
This work was supported by FCT - Fundação para a Ciência e a Tecnologia through national funds and by FEDER through COMPETE2020 - Programa Operacional Competitividade e Internacionalizacão by these grants: UID/FIS/04434/2019, UIDB/04434/2020, UIDP/04434/2020, PTDC/FIS-AST/32113/2017 \& POCI-01-0145-FEDER- 032113, PTDC/FIS-AST/28953/2017 \& POCI-01-0145-FEDER-028953, PTDC/FIS-AST/28987/2017 \& POCI-01-0145-FEDER-028987, O.D.S.D. is supported in the form of work contract (DL 57/2016/CP1364/CT0004) funded by national funds through FCT.\\ %Olivier D. S. Demangeon
We acknowledge support from the Spanish Ministry of Science and Innovation and the European Regional Development Fund through grants ESP2016-80435-C2-1-R, ESP2016-80435-C2-2-R, PGC2018-098153-B-C33, PGC2018-098153-B-C31, ESP2017-87676-C5-1-R, MDM-2017-0737 Unidad de Excelencia Maria de Maeztu-Centro de Astrobiolog\'{i}a (INTA-CSIC), as well as the support of the Generalitat de Catalunya/CERCA programme. The MOC activities have been supported by the ESA contract No. 4000124370.\\ %Roi Alonso
XB, SC, DG, MF and JL acknowledge their role as ESA-appointed CHEOPS science team members.\\ %Xavier Bonfils
This project has received funding from the European Research Council (ERC) under the European Union’s Horizon 2020 research and innovation programme (project {\sc Four Aces}; grant agreement No 724427); DE acknowledges financial support from the Swiss National Science Foundation for project 200021\_200726.\\ %David Ehrenreich
(covered by Tom Wilson's acknowledgement)\\ %Andrew Collier Cameron
This work was also partially supported by a grant from the Simons Foundation (PI Queloz, grant number 327127).\\ %Didier Queloz
SH gratefully acknowledges CNES funding through the grant 837319.\\ %Sergio Hoyer
KGI is the ESA CHEOPS Project Scientist and is responsible for the ESA CHEOPS Guest Observers Programme. She does not participate in, or contribute to, the definition of the Guaranteed Time Programme of the CHEOPS mission through which observations described in this paper have been taken, nor to any aspect of target selection for the programme.\\ %Kate G. Isaak
YGMC is supported by UNAM-PAPIIT-IG101321. This work includes observations with SAINT-EX carried out at the Observatorio Astronómico Nacional on the Sierra de San Pedro M\'artir (OAN-SPM), Baja California, M\'exico.\\ %Y. Gómez~Maqueo~Chew
S.S. have received funding from the European Research Council (ERC) under the European Unions Horizon 2020 research and innovation programme (grant agreement No 833925,  project STAREX).\\ %Sébastien Salmon
\end{acknowledgements}

%-------------------------------------------------------------------
\bibliographystyle{aa}
\bibliography{refs.bib}

\begin{appendix}

\section{TTV modelling}\label{app:ttv}
\refcom{We performed three TTV modelling approaches to derive planetary parameters and assess how prior-dependent these models are.
In the first approach, we used the \texttt{TTVfaster} package \citep{2016ascl.soft04012A} to generate models of TTVs given input parameters for the three planets \& star.
\texttt{TTVfaster} requires the assumption that the planets are not in perfect resonant orbits with one another.
Using the periods and epochs, we find that this assumption appears to be satisfied for b \& c ($P_c/P_b = 2.0296$), but we cannot be sure about c \& d, which are closer to a resonant ratio ($P_d/P_c = 1.6713 = 5.0140/3$).}
%\textbf{[Adrien to complete resonance check]}

\refcom{As we had many parameters and few transit times with which to constrain them (which could result in multi-modal parameter space), we used a nested sampling approach which is better able to explore non-Gaussian parameter space than a simple MCMC \citep{2021arXiv210109675B}.
We used \texttt{Ultranest} for this implementation \citep{2021JOSS....6.3001B}, and used the \texttt{stepsampler.RegionSliceSampler} method as the number of parameters is large.}

\refcom{For period and inclination priors, we used outputs from our combined model as Gaussian priors, but increased the standard deviation by a factor of 2.5 to limit any over-fitting.
The best-fit transit epoch from the combined model was used as a normal prior, with a standard deviation of 0.05\,d (far larger than the timing fit uncertainty to prevent overfitting).
The longitude of ascending node and an argument of periastron were given wide uniform priors from -$\pi$ to $\pi$.
For eccentricity we used the half-normal distribution of multi-planet systems from \citet{2015ApJ...808..126V} ($\sigma = 0.096$).
Although typical samplers such as MCMC struggle due to correlations when not exploring $e\cos{\omega}$ \& $e\sin{\omega}$, we found this had little effect on our nested sampling results, likely because samples are independent from their predecessors.
For the outer planets we reparameterised planetary masses as log mass ratios and planetary periods as simple ratios to avoid strong correlations (for planet b, as a ratio to the star, and for planets c \& d as a ratio of planet b).
For planetary mass ratios, we used the population of exoplanets with well-constrained masses and radii \citep[Downloaded from the NASA exoplanet archive, ][]{2013PASP..125..989A} to produce broad Gaussian priors on log planetary mass ($\log{M_p}$) given a planetary radius.
This resulted in mass priors of $7.8^{+4.3}_{-2.8}$, $11.3^{+7.7}_{-4.6}$ and $10.2^{+6.7}_{-4.0} M_\oplus$ for planets b, c \& d respectively, \refcom{which match very closely the predictions of \texttt{forecaster}\citep{chen2016probabilistic}}. 
We inflated these standard deviations from the $\log{M_p}$ population prior by 0.1 to prevent overly constraining priors.}
%From forecaster: $7.2^{+5.4}_{-3.2}$, $12.3^{+9.7}_{-5.4}$ and $10.8^{+7.5}_{-4.5}$.

\refcom{Indepedently, and in an effort to estimate the influence of the mass and eccentricity priors on the determined posterior -- and to take into account the possible resonant motion of the outer pair ($c$ and $d$ are very close to the exact commensurability: $P_d/P_c-5/3=0.0047$) -- we use the approach presented in \cite{2021A&A...655A..66L}.
Here we estimated transit timing variations are estimated using the {\ttfamily TTVfast} algorithm \citep{ttvfast2014}, and the \texttt{samsam}\footnote{\url{https://gitlab.unige.ch/Jean-Baptiste.Delisle/samsam}.} MCMC algorithm \citep[see][]{Delisle2018} is used to sample the posterior.
Following \citep{HaLi2017}, we test the robustness of TTV mass-estimation by trying out two mass priors: log10-uniform and uniform. 
The mass and eccentricity posteriors are shown in Table \ref{tab:ttvres}.}

\refcom{As shown in Table \ref{tab:ttvres}, the determined masses depend strongly on the used priors.
The nested sampling approach appears to find low but plausible masses for all three planets: $M_b=$\TttvMb{}, $M_c=$\TttvMc{}, and $M_d=$\TttvMd{}$M_\oplus$. 
Best-fit TTV models from this approach are shown in Fig~\ref{fig:TTVmodels}.
However planets c \& d may be in 5:3 resonance, in which case the models of \texttt{TTVFaster} are not valid.
In addition, the inner pair is close, but not inside, a mean motion resonance, which creates degeneracy between the determined masses and eccentricities \citep{LiWu2012}.
The \cite{2021A&A...655A..66L} approach finds extremely small ($M_p<1M_\oplus$) and large ($M_p>10M_\oplus$) for the log-uniform \& uniform mass priors respectively.}

%, and therefore the data is not sufficient to estimate the masses from TTVs.
%All model priors and posteriors are shown in Table \ref{table:appendix_ttv_model}.
%As we have relatively few transits, the TTV model did not converge onto a single local minimum for all planets and across our three models. %, and there remains a large range of possible models which could fit the data (see 

%Using the radii from our combined photometric model and the masses from our TTV model, we can produce the first density estimates for the planets around TOI-2076. 
%We find that, with densities of \Tttvrhob{}, \Tttvrhoc{} \& \Tttvrhod{}\gccc{}, the three mini-Neptunes appear to have low densities and therefore likely contain thick gas layers.
%The densities also hint that TOI-2076~b follows the trend of inner planets in systems of multiple Neptunes having a higher than average density, suggesting it may have undergone significant evaporation.

%We do not expect as large an observable TTV amplitude between planets c and d, as the 5:3 resonance is inherently weaker than the first-order 2:1, and the proximity of the period to the resonance (0.3\%) means any super-period is likely far longer. 
%However, the so-called "chopping" signal, which act close to the orbital period of the inner planet may provide stochastic noise in our timing measurements.
\begin{table}
%\centering                          % used for centering table
\caption{Priors and posteriors for planetary masses and eccentricities from each of the three TTV models used.}             % title of Table
\label{tab:ttvres}      % is used to refer this table in the text
\begin{tabular}{l r c | c | c }        % centred columns (4 columns)
\hline\hline                 % inserts double horizontal lines
Param. & Prior Type & pl & Prior & Posterior \\
\hline
\multicolumn{3}{l|}{\textbf{TTVFaster+Nested Sampling}} & & \\
Mass [$M_\oplus$] & log-Normal & \multirow{2}{*}{b} &  $7.7^{+5.6}_{-3.2}$ & \TttvMb{} \\
Eccentricity & half-Normal  & & $0.0^{+0.096}_{-0}$ & \Teccb{} \\
Mass [$M_\oplus$] & log-Normal & \multirow{2}{*}{c} &  $11.3^{+9.7}_{-5.2}$ & \TttvMc{} \\
Eccentricity & half-Normal  & & $0.0^{+0.096}_{-0}$ & \Teccc{} \\
Mass [$M_\oplus$] & log-Normal & \multirow{2}{*}{d} &  $10.2^{+8.3}_{-4.6}$ & \TttvMd{} \\
Eccentricity & half-Normal  & & $0.0^{+0.096}_{-0}$ & \Teccd{} \\
\hline                        % inserts single horizontal line
\multicolumn{3}{l|}{\textbf{N-body+MCMC}} & & \\
Mass [$M_\oplus$] & log-Uniform & \multirow{2}{*}{b} &  $ [.03,3000]$ & $0.62 \pm 0.50$ \\
Eccentricity & Uniform  & & $ [0,0.9] $ & $0.204 \pm 0.099 $ \\
Mass [$M_\oplus$] & log-Uniform & \multirow{2}{*}{c} &  [.03,3000] & $0.84 \pm 0.58$ \\
Eccentricity & Uniform & & $[0,0.9] $ & $0.038 \pm 0.029 $ \\
Mass [$M_\oplus$] & log-Uniform & \multirow{2}{*}{d} &  [.03,3000] & $0.74 \pm 1.07$ \\
Eccentricity & Uniform & & $[0,0.9] $ & $ 0.037 \pm 0.026$ \\
\hline
\multicolumn{3}{l|}{\textbf{N-body+MCMC}} & & \\
Mass [$M_\oplus$] & Uniform & \multirow{2}{*}{b} &  [.03,3000] & $45.68 \pm 22.09$ \\
Eccentricity & Uniform & & $ [0,0.9] $ & $0.0042 \pm 0.0033 $ \\
Mass [$M_\oplus$] & Uniform & \multirow{2}{*}{c} &  [.03,3000] & $12.43 \pm 2.72$ \\ %Place stupid priors; win stupid posteriors
Eccentricity & Uniform & & $[0,0.9] $ & $0.0079 \pm 0.0066 $ \\
Mass [$M_\oplus$] & Uniform & \multirow{2}{*}{d} &  [.03,3000] & $ 97.98 \pm 60.01$ \\
Eccentricity & Uniform & & $[0,0.9] $ & $0.0072  \pm 0.0059 $ \\
\hline                                   %inserts single line
\end{tabular}
\end{table}

\section{Combined model parameters}
\begin{table*}
\tiny
\caption{Model parameters, priors, and posteriors for the Combined model.} 
\label{table:appendix_P35_model}      % is used to refer this table in the text
\centering                          % used for centering table
\begin{tabular}{l c c}        % centred columns (4 columns)
\hline\hline                 % inserts double horizontal lines
Parameter & Prior & Posterior \\
\hline                        % inserts single horizontal line
Stellar temperature, \teff{}  [K]  & $\mathcal{N}_{\mathcal{U}}(a=4000,b=6000,\mu=5200,\sigma=68)$ &  $ 5200.0 \pm 66.0 $  \\
Stellar radius, $R_s$ [$R_\oplus$] & $\mathcal{N}_{\mathcal{U}}(a=0,\mu=0.77,\sigma=0.006)$ &  $ 0.7699 \pm 0.0059 $  \\
log stellar surface gravity, \logg{}  [cgs] & $\mathcal{N}(\mu=4.45,\sigma=0.12) $ &  $ 4.576^{+0.012}_{-0.017} $  \\
Transit time, $t_{b,0}$ [BJD-2457000] & $\mathcal{N}(\mu=1743.73,\sigma=0.025)$ &  $ 1743.7193 \pm 0.0022 $  \\
Transit time, $t_{b,1}$ [BJD-2457000] & $\mathcal{N}(\mu=1754.08,\sigma=0.025)$ &  $ 1754.0776 \pm 0.0012 $  \\
Transit time, $t_{b,2}$ [BJD-2457000] & $\mathcal{N}(\mu=1930.12,\sigma=0.025)$ &  $ 1930.1221 \pm 0.002 $  \\
Transit time, $t_{b,3}$ [BJD-2457000] & $\mathcal{N}(\mu=1940.47,\sigma=0.025)$ &  $ 1940.4798 \pm 0.0011 $  \\
Transit time, $t_{b,4}$ [BJD-2457000] & $\mathcal{N}(\mu=1950.83,\sigma=0.025)$ &  $ 1950.8343 \pm 0.0013 $  \\
Transit time, $t_{b,5}$ [BJD-2457000] & $\mathcal{N}(\mu=2333.96,\sigma=0.025)$ &  $ 2333.9547 \pm 0.0024 $  \\
Transit time, $t_{c,0}$ [BJD-2457000] & $\mathcal{N}(\mu=1748.69,\sigma=0.025)$ &  $ 1748.69408 \pm 0.00079 $  \\
Transit time, $t_{c,1}$ [BJD-2457000] & $\mathcal{N}(\mu=1937.83,\sigma=0.025)$ &  $ 1937.82201 \pm 0.0008 $  \\
Transit time, $t_{c,2}$ [BJD-2457000] & $\mathcal{N}(\mu=2274.08,\sigma=0.025)$ &  $ 2274.08398 \pm 0.00079 $  \\
Transit time, $t_{d,0}$ [BJD-2457000] & $\mathcal{N}(\mu=1762.67,\sigma=0.025)$ &  $ 1762.6679 \pm 0.0016 $  \\
Transit time, $t_{d,1}$ [BJD-2457000] & $\mathcal{N}(\mu=1938.29,\sigma=0.025)$ &  $ 1938.2915 \pm 0.0014 $  \\
Transit time, $t_{d,2}$ [BJD-2457000] & $\mathcal{N}(\mu=2359.79,\sigma=0.025)$ &  $ 2359.789 \pm 0.022 $  \\
Transit time, $t_{d,3}$ [BJD-2457000] & $\mathcal{N}(\mu=2394.91,\sigma=0.025)$ &  $ 2394.9236 \pm 0.0015 $  \\
Log radius ratio, $\log{R_{p,b}/R_s}$  & $\mathcal{N}(\mu=-3.48794,\sigma=1)$ &  $ -3.507 \pm 0.012 $  \\
Log radius ratio, $\log{R_{p,c}/R_s}$  & $\mathcal{N}(\mu=-3.18726,\sigma=1)$ &  $ -3.1788^{+0.0093}_{-0.0098} $  \\
Log radius ratio, $\log{R_{p,d}/R_s}$  & $\mathcal{N}(\mu=-3.14438,\sigma=1)$ &  $ -3.258 \pm 0.018 $  \\
Impact parameter, $b_{0}$  & $ \mathcal{{U}}(a=0.0,b=1+R_{p,b}/R_s)^\ddagger{}$ &  $ 0.149 \pm 0.089 $  \\
Impact parameter, $b_{1}$  & $ \mathcal{{U}}(a=0.0,b=1+R_{p,c}/R_s)^\ddagger{}$ &  $ 0.092^{+0.092}_{-0.063} $  \\
Impact parameter, $b_{2}$  & $ \mathcal{{U}}(a=0.0,b=1+R_{p,d}/R_s)^\ddagger{}$ &  $ 0.8225 \pm 0.0087 $  \\
Quadratic LD, $u_{{\rm cheops},0} $  & $\mathcal{N}_{\mathcal{U}}(a=0.5015,b=0.5707,\mu=0.5367,\sigma=0.0500)$ &  $ 0.567 \pm 0.038 $  \\
Quadratic LD, $u_{{\rm cheops},1} $  & $\mathcal{N}_{\mathcal{U}}(a=0.1457,b=0.1949,\mu=0.1705,\sigma=0.0500)$ &  $ 0.187 \pm 0.047 $  \\
Quadratic LD, $u_{{\rm g},0} $  & $\mathcal{N}_{\mathcal{U}}(a=0.6800,b=0.7732,\mu=0.7257,\sigma=0.0500)$ &  $ 0.701 \pm 0.048 $  \\
Quadratic LD, $u_{{\rm g},1} $  & $\mathcal{N}_{\mathcal{U}}(a=0.0513,b=0.1269,\mu=0.0911,\sigma=0.0500)$ &  $ 0.081^{+0.047}_{-0.044} $  \\
Quadratic LD, $u_{{\rm i},0} $  & $\mathcal{N}_{\mathcal{U}}(a=0.3776,b=0.4283,\mu=0.4043,\sigma=0.0500)$ &  $ 0.389 \pm 0.049 $  \\
Quadratic LD, $u_{{\rm i},1} $  & $\mathcal{N}_{\mathcal{U}}(a=0.2043,b=0.2355,\mu=0.2186,\sigma=0.0500)$ &  $ 0.206 \pm 0.05 $  \\
Quadratic LD, $u_{{\rm r},0} $  & $\mathcal{N}_{\mathcal{U}}(a=0.4771,b=0.5458,\mu=0.5114,\sigma=0.0500)$ &  $ 0.477 \pm 0.046 $  \\
Quadratic LD, $u_{{\rm r},1} $  & $\mathcal{N}_{\mathcal{U}}(a=0.1800,b=0.2255,\mu=0.2025,\sigma=0.0500)$ &  $ 0.182 \pm 0.048 $  \\
Quadratic LD, $u_{{\rm tess},0} $  & $\mathcal{N}_{\mathcal{U}}(a=0.3703,b=0.4255,\mu=0.3981,\sigma=0.0500)$ &  $ 0.375 \pm 0.04 $  \\
Quadratic LD, $u_{{\rm tess},1} $  & $\mathcal{N}_{\mathcal{U}}(a=0.2046,b=0.2383,\mu=0.2219,\sigma=0.0500)$ &  $ 0.208 \pm 0.046 $  \\
Quadratic LD, $u_{{\rm z},0} $  & $\mathcal{N}_{\mathcal{U}}(a=0.2028,b=0.3076,\mu=0.2333,\sigma=0.0500)$ &  $ 0.212 \pm 0.048 $  \\
Quadratic LD, $u_{{\rm z},1} $  & $\mathcal{N}_{\mathcal{U}}(a=0.2428,b=0.3645,\mu=0.3251,\sigma=0.0500)$ &  $ 0.31 \pm 0.05 $  \\
Log photometric scatter, $\log{\sigma_{{\rm g_lco},s}/(\rm{ppt})}$  & $\mathcal{N}(\mu=3.535,\sigma=3) $ &  $ -3.3 \pm 1.0 $  \\
Log photometric scatter, $\log{\sigma_{{\rm r_lco},s}/(\rm{ppt})}$  & $\mathcal{N}(\mu=2.947,\sigma=3) $ &  $ -3.6 \pm 1.1 $  \\
Log photometric scatter, $\log{\sigma_{{\rm i_lco},s}/(\rm{ppt})}$  & $\mathcal{N}(\mu=2.557,\sigma=3) $ &  $ -3.6 \pm 1.1 $  \\
Log photometric scatter, $\log{\sigma_{{\rm z_lco},s}/(\rm{ppt})}$  & $\mathcal{N}(\mu=2,\sigma=3) $ &  $ -3.6 \pm 1.2 $  \\
Log photometric scatter, $\log{\sigma_{{\rm r_sex},s}/(\rm{ppt})}$  & $\mathcal{N}(\mu=2.101,\sigma=3) $ &  $ -1.5^{+1.2}_{-1.7} $  \\
Log photometric scatter, $\log{\sigma_{{\rm z_mcd},s}/(\rm{ppt})}$  & $\mathcal{N}(\mu=0.9765,\sigma=3) $ &  $ 0.4^{+2.1}_{-2.7} $  \\
Log photometric scatter, $\log{\sigma_{{\rm cheops_0},s}/(\rm{ppt})}$  & $\mathcal{N}(\mu=-0.7551,\sigma=3) $ &  $ -1.816 \pm 0.09 $  \\
Log photometric scatter, $\log{\sigma_{{\rm cheops_1},s}/(\rm{ppt})}$  & $\mathcal{N}(\mu=-0.5174,\sigma=3) $ &  $ -1.85 \pm 0.1 $  \\
Log photometric scatter, $\log{\sigma_{{\rm cheops_2},s}/(\rm{ppt})}$  & $\mathcal{N}(\mu=-0.3489,\sigma=3) $ &  $ -1.221 \pm 0.049 $  \\
Log photometric scatter, $\log{\sigma_{{\rm cheops_3},s}/(\rm{ppt})}$  & $\mathcal{N}(\mu=0.5838,\sigma=3) $ &  $ -0.991 \pm 0.048 $  \\
Log photometric scatter, $\log{\sigma_{{\rm tess},s}/(\rm{ppt})}$  & $\mathcal{N}(\mu=-0.314,\sigma=3) $ &  $ -1.338 \pm 0.037 $  \\
g-lco airmass trend, $df/d({\rm airmass})_N$  & $ \mathcal{{U}}(a=-35,b=10) $ &  $ -22.87 \pm 0.91 $  \\
g-lco aperture entropy trend, $df/d({\rm entropy})_N$  & $\mathcal{{N}}_{{\mathcal{{U}}}}(a=-20,b=20,\mu=0,\sigma=1)$ &  $ 0.11 \pm 0.24 $  \\
g-lco time trend, $df/d({\rm time})_N$  & $\mathcal{{N}}_{{\mathcal{{U}}}}(a=-20,b=20,\mu=0,\sigma=1)$ &  $ -0.9 \pm 0.26 $  \\
g-lco aperture width trend, $df/d({\rm width})_N$  & $\mathcal{{N}}_{{\mathcal{{U}}}}(a=-20,b=20,\mu=0,\sigma=1)$ &  $ -0.65 \pm 0.18 $  \\
g-lco g/r colour trend, $df/d({\rm g/r})_N$  & $\mathcal{{N}}_{{\mathcal{{U}}}}(a=-20,b=20,\mu=0,\sigma=1)$ &  $ 7.55 \pm 0.67 $  \\
g-lco r/i colour trend, $df/d({\rm r/i})_N$  & $\mathcal{{N}}_{{\mathcal{{U}}}}(a=-20,b=20,\mu=0,\sigma=1)$ &  $ 3.41 \pm 0.19 $  \\
g-lco airmass quadratic, $d^2f/d({\rm airmass})^2_N$  & $ \mathcal{{U}}(a=-35,b=10) $ &  $ 0.68 \pm 0.11 $  \\
r-lco airmass trend, $df/d({\rm airmass})_N$  & $ \mathcal{{U}}(a=-35,b=10) $ &  $ -15.94 \pm 0.73 $  \\
r-lco time trend, $df/d({\rm time})_N$  & $\mathcal{{N}}_{{\mathcal{{U}}}}(a=-20,b=20,\mu=0,\sigma=1)$ &  $ -1.31 \pm 0.18 $  \\
r-lco aperture width trend, $df/d({\rm width})_N$  & $\mathcal{{N}}_{{\mathcal{{U}}}}(a=-20,b=20,\mu=0,\sigma=1)$ &  $ -0.438 \pm 0.059 $  \\
r-lco g/r colour trend, $df/d({\rm g/r})_N$  & $\mathcal{{N}}_{{\mathcal{{U}}}}(a=-20,b=20,\mu=0,\sigma=1)$ &  $ -1.35 \pm 0.59 $  \\
r-lco r/i colour trend, $df/d({\rm r/i})_N$  & $\mathcal{{N}}_{{\mathcal{{U}}}}(a=-20,b=20,\mu=0,\sigma=1)$ &  $ 2.76 \pm 0.16 $  \\
r-lco airmass quadratic, $d^2f/d({\rm airmass})^2_N$  & $ \mathcal{{U}}(a=-35,b=10) $ &  $ 0.163 \pm 0.082 $  \\

\hline                                   %inserts single line
\end{tabular}
    \begin{tablenotes}
      \tiny
      \item $\mathcal{N}$ details a normally distributed prior with mean, $\mu$ and standard deviation, $\sigma$ values. $\mathcal{U}$ details a uniform distribution with lower, $a$, and upper, $b$, limits. $\mathcal{N}_\mathcal{U}$ details a truncated normal distribution with $\mu$,$\sigma$, $a$ \& $b$ values.$\ddagger{}$ represents the uniform prior as presented by \citet{2018RNAAS...2..209E} and implemented by \texttt{exoplanet}. \cheops{} suffixes refer chronologically to the four unique \cheops{} visits, SaEx refers to detrending parameters for the photometry from SAINT-EX, McD refers to those for photometry from the 1m LCO telescope at McDonald, and lco refers to data from the 2m LCO telescope with the MuSCAT-3 instrument in each of the four bands (g-, r-, i-, \& z-).
    \end{tablenotes}

\end{table*}

\begin{table*}
\tiny
\caption{Model parameters, priors and posteriors for the Combined model (Continued from Table \ref{table:appendix_P35_model})}             % title of Table
\label{table:appendix_P35_model2}      % is used to refer this table in the text
\centering                          % used for centering table
\begin{tabular}{l c c}        % centred columns (4 columns)
\hline\hline                 % inserts double horizontal lines
Parameter & Prior & Posterior \\
\hline                        % inserts single horizontal line
i-lco airmass trend, $df/d({\rm airmass})_N$  & $ \mathcal{{U}}(a=-35,b=10) $ &  $ -11.24^{+0.96}_{-0.93} $  \\
i-lco aperture entropy trend, $df/d({\rm entropy})_N$  & $\mathcal{{N}}_{{\mathcal{{U}}}}(a=-20,b=20,\mu=0,\sigma=1)$ &  $ 0.16 \pm 0.15 $  \\
i-lco time trend, $df/d({\rm time})_N$  & $\mathcal{{N}}_{{\mathcal{{U}}}}(a=-20,b=20,\mu=0,\sigma=1)$ &  $ -2.38 \pm 0.33 $  \\
i-lco aperture width trend, $df/d({\rm width})_N$  & $\mathcal{{N}}_{{\mathcal{{U}}}}(a=-20,b=20,\mu=0,\sigma=1)$ &  $ -0.47 \pm 0.17 $  \\
i-lco companion Flux trend, $df/dF_{{\rm comps},N}$  & $\mathcal{{N}}_{{\mathcal{{U}}}}(a=-20,b=20,\mu=0,\sigma=1)$ &  $ 0.48 \pm 0.19 $  \\
i-lco r/i colour trend, $df/d({\rm r/i})_N$  & $\mathcal{{N}}_{{\mathcal{{U}}}}(a=-20,b=20,\mu=0,\sigma=1)$ &  $ -3.69 \pm 0.33 $  \\
i-lco i/z colour trend, $df/d({\rm i/z})_N$  & $\mathcal{{N}}_{{\mathcal{{U}}}}(a=-20,b=20,\mu=0,\sigma=1)$ &  $ 1.01 \pm 0.4 $  \\
i-lco airmass quadratic, $d^2f/d({\rm airmass})^2_N$  & $ \mathcal{{U}}(a=-35,b=10) $ &  $ -0.39 \pm 0.1 $  \\
z-lco airmass trend, $df/d({\rm airmass})_N$  & $ \mathcal{{U}}(a=-35,b=10) $ &  $ -8.8 \pm 1.0 $  \\
z-lco time trend, $df/d({\rm time})_N$  & $\mathcal{{N}}_{{\mathcal{{U}}}}(a=-20,b=20,\mu=0,\sigma=1)$ &  $ -2.38 \pm 0.3 $  \\
z-lco aperture width trend, $df/d({\rm width})_N$  & $\mathcal{{N}}_{{\mathcal{{U}}}}(a=-20,b=20,\mu=0,\sigma=1)$ &  $ -0.3 \pm 0.12 $  \\
z-lco r/i colour trend, $df/d({\rm r/i})_N$  & $\mathcal{{N}}_{{\mathcal{{U}}}}(a=-20,b=20,\mu=0,\sigma=1)$ &  $ -2.32 \pm 0.42 $  \\
z-lco i/z colour trend, $df/d({\rm i/z})_N$  & $\mathcal{{N}}_{{\mathcal{{U}}}}(a=-20,b=20,\mu=0,\sigma=1)$ &  $ -3.21 \pm 0.48 $  \\
z-lco airmass quadratic, $d^2f/d({\rm airmass})^2_N$  & $ \mathcal{{U}}(a=-35,b=10) $ &  $ -0.48 \pm 0.12 $  \\
r-SaEx airmass trend, $df/d({\rm airmass})_N$  & $ \mathcal{{U}}(a=-35,b=10) $ &  $ 2.3 \pm 1.3 $  \\
r-SaEx companion Flux trend, $df/dF_{{\rm comps},N}$  & $\mathcal{{N}}_{{\mathcal{{U}}}}(a=-20,b=20,\mu=0,\sigma=1)$ &  $ -3.87 \pm 0.77 $  \\
r-SaEx airmass quadratic, $d^2f/d({\rm airmass})^2_N$  & $ \mathcal{{U}}(a=-35,b=10) $ &  $ -0.82 \pm 0.63 $  \\
z-McD airmass trend, $df/d({\rm airmass})_N$  & $\mathcal{{N}}_{{\mathcal{{U}}}}(a=-20,b=20,\mu=0,\sigma=1)$ &  $ 0.0 \pm 1.0 $  \\
Cheops-0 time trend, $df/d({\rm time})_N$  & $\mathcal{{N}}_{{\mathcal{{U}}}}(a=-20,b=20,\mu=0,\sigma=1)$ &  $ 0.358 \pm 0.012 $  \\
Cheops-0 cosine of Rollangle slope, $df/d({\cos{\Phi}})_N$  & $\mathcal{{N}}_{{\mathcal{{U}}}}(a=-20,b=20,\mu=0,\sigma=1)$ &  $ -0.061 \pm 0.014 $  \\
Cheops-0 background flux slope, $df/d{\rm bg})_N$  & $\mathcal{{N}}_{{\mathcal{{U}}}}(a=-20,b=20,\mu=0,\sigma=1)$ &  $ 0.146 \pm 0.014 $  \\
Cheops-0 time quadratic, $d^2f/d({\rm time})^2_N$  & $\mathcal{{N}}_{{\mathcal{{U}}}}(a=-20,b=20,\mu=0,\sigma=1)$ &  $ -0.057 \pm 0.013 $  \\
Cheops-1 time trend, $df/d({\rm time})_N$  & $\mathcal{{N}}_{{\mathcal{{U}}}}(a=-20,b=20,\mu=0,\sigma=1)$ &  $ 0.056^{+0.014}_{-0.015} $  \\
Cheops-1 sine of Rollangle slope, $df/d({\sin{\Phi}})_N$   & $\mathcal{{N}}_{{\mathcal{{U}}}}(a=-20,b=20,\mu=0,\sigma=1)$ &  $ -0.03 \pm 0.013 $  \\
Cheops-1 cosine of Rollangle slope, $df/d({\cos{\Phi}})_N$  & $\mathcal{{N}}_{{\mathcal{{U}}}}(a=-20,b=20,\mu=0,\sigma=1)$ &  $ -0.013 \pm 0.014 $  \\
Cheops-1 CCD smear slope, $df/d{\rm smear})_N$  & $\mathcal{{N}}_{{\mathcal{{U}}}}(a=-20,b=20,\mu=0,\sigma=1)$ &  $ 0.013 \pm 0.012 $  \\
Cheops-1 background flux slope, $df/d{\rm bg})_N$  & $\mathcal{{N}}_{{\mathcal{{U}}}}(a=-20,b=20,\mu=0,\sigma=1)$ &  $ 0.097 \pm 0.014 $  \\
Cheops-1 time quadratic, $d^2f/d({\rm time})^2_N$  & $\mathcal{{N}}_{{\mathcal{{U}}}}(a=-20,b=20,\mu=0,\sigma=1)$ &  $ -0.093^{+0.015}_{-0.015} $  \\
Cheops-2 time trend, $df/d({\rm time})_N$  & $\mathcal{{N}}_{{\mathcal{{U}}}}(a=-20,b=20,\mu=0,\sigma=1)$ &  $ 0.591 \pm 0.016 $  \\
Cheops-2 sine of Rollangle slope, $df/d({\sin{\Phi}})_N$   & $\mathcal{{N}}_{{\mathcal{{U}}}}(a=-20,b=20,\mu=0,\sigma=1)$ &  $ -0.019 \pm 0.016 $  \\
Cheops-2 cosine of Rollangle slope, $df/d({\cos{\Phi}})_N$  & $\mathcal{{N}}_{{\mathcal{{U}}}}(a=-20,b=20,\mu=0,\sigma=1)$ &  $ 0.006 \pm 0.017 $  \\
Cheops-2 background flux slope, $df/d{\rm bg})_N$  & $\mathcal{{N}}_{{\mathcal{{U}}}}(a=-20,b=20,\mu=0,\sigma=1)$ &  $ 0.08 \pm 0.017 $  \\
Cheops-2 time quadratic, $d^2f/d({\rm time})^2_N$  & $\mathcal{{N}}_{{\mathcal{{U}}}}(a=-20,b=20,\mu=0,\sigma=1)$ &  $ -0.012 \pm 0.017 $  \\
Cheops-3 time trend, $df/d({\rm time})_N$  & $\mathcal{{N}}_{{\mathcal{{U}}}}(a=-20,b=20,\mu=0,\sigma=1)$ &  $ -1.243 \pm 0.024 $  \\
Cheops-3 cosine of Rollangle slope, $df/d({\cos{\Phi}})_N$  & $\mathcal{{N}}_{{\mathcal{{U}}}}(a=-20,b=20,\mu=0,\sigma=1)$ &  $ 0.025 \pm 0.023 $  \\
Cheops-3 CCD smear slope, $df/d{\rm smear})_N$  & $\mathcal{{N}}_{{\mathcal{{U}}}}(a=-20,b=20,\mu=0,\sigma=1)$ &  $ -0.027 \pm 0.021 $  \\
Cheops-3 background flux slope, $df/d{\rm bg})_N$  & $\mathcal{{N}}_{{\mathcal{{U}}}}(a=-20,b=20,\mu=0,\sigma=1)$ &  $ 0.015 \pm 0.022 $  \\
Cheops-3 time quadratic, $d^2f/d({\rm time})^2_N$  & $\mathcal{{N}}_{{\mathcal{{U}}}}(a=-20,b=20,\mu=0,\sigma=1)$ &  $ 0.136 \pm 0.028 $  \\
Mean flux, $\mu_{{\rm g-lco}}$ [ppt]  & $\mathcal{N}(\mu=0,\sigma=17.15) $ &  $ 0.24 \pm 0.054 $  \\
Mean flux, $\mu_{{\rm r-lco}}$ [ppt]  & $\mathcal{N}(\mu=0,\sigma=9.529) $ &  $ 0.469 \pm 0.049 $  \\
Mean flux, $\mu_{{\rm i-lco}}$ [ppt]  & $\mathcal{N}(\mu=0,\sigma=6.451) $ &  $ 0.232 \pm 0.07 $  \\
Mean flux, $\mu_{{\rm z-lco}}$ [ppt]  & $\mathcal{N}(\mu=0,\sigma=3.694) $ &  $ 0.385 \pm 0.092 $  \\
Mean flux, $\mu_{{\rm r-SaEx}}$ [ppt]  & $\mathcal{N}(\mu=0,\sigma=4.088) $ &  $ 0.89 \pm 0.92 $  \\
Mean flux, $\mu_{{\rm z-McD}}$ [ppt]  & $\mathcal{N}(\mu=0,\sigma=1.328) $ &  $ 0.0 \pm 1.3 $  \\
Mean flux, $\mu_{{\rm Cheops-0}}$ [ppt]  & $\mathcal{N}(\mu=0,\sigma=0.235) $ &  $ -0.033 \pm 0.018 $  \\
Mean flux, $\mu_{{\rm Cheops-1}}$ [ppt]  & $\mathcal{N}(\mu=0,\sigma=0.298) $ &  $ 0.239 \pm 0.022 $  \\
Mean flux, $\mu_{{\rm Cheops-2}}$ [ppt]  & $\mathcal{N}(\mu=0,\sigma=0.3527) $ &  $ 0.016 \pm 0.024 $  \\
Mean flux, $\mu_{{\rm Cheops-3}}$ [ppt]  & $\mathcal{N}(\mu=0,\sigma=0.8964) $ &  $ 1.295^{+0.045}_{-0.045} $  \\

\hline                                   %inserts single line
\end{tabular}
\end{table*}

\section{TTVFaster model parameters}
\begin{table*}
\caption{Model parameters, priors and posteriors for the \texttt{TTVfaster/Ultranest} TTV model.}             % title of Table
\label{table:appendix_ttv_model}      % is used to refer this table in the text
\centering                          % used for centering table
\begin{tabular}{l c c}        % centred columns (4 columns)
\hline\hline                 % inserts double horizontal lines
Parameter & Prior & Posterior \\
\hline                        % inserts single horizontal line
Stellar Mass $ M_s $ [$M_\odot$] &  $\mathcal{N}(\mu=0.865,\sigma=0.036)$ &  $ 0.882^{+0.028}_{-0.051} $  \\
log mass ratio, $ \log{M_{p,b} / M_{s}} $ &  $\mathcal{N}(\mu=-10.52,\sigma=0.58)$ &  $ -10.82^{+0.39}_{-0.6} $  \\
Period, $P_b$ [d] &  $\mathcal{N}_{\mathcal{U}}(a=10.325,b=10.385,\mu=10.355,\sigma=0.0013)$ &  $ 10.35509^{+0.0002}_{-0.00014} $  \\
$ e_{b} $ &  $|\mathcal{N}(0,0.096)|$ &  $ 0.023 \pm 0.02 $  \\
$ \omega_{b} $ &  $\mathcal{U}(a=-3.14,b=3.14)$ &  $ -0.6 \pm 1.7 $  \\
Inclination, $ i_{b} $ [$^{\circ}$] &  $\mathcal{N}(\mu=1.563,\sigma=0.011)$ &  $ 1.567 \pm 0.013 $  \\
Longitude of Ascending Node, $ \Omega_{b} $ [$^{\circ}$] &  $\mathcal{U}(a=-3.14,b=3.14)$ &  $ 0.2 \pm 1.9 $  \\
Transit Epoch, $ t_{0, b} $ [BJD-2457000] &  $\mathcal{N}(\mu=1743.728,\sigma=0.05)$ &  $ 1743.7231^{+0.0061}_{-0.0087} $  \\
log mass ratio, $ \log{M_{p,c} / M_{p, b}} $ &  $\mathcal{N}_{\mathcal{U}}(a=-2.62,b=3.38,\mu=0.38,\sigma=0.62)$ &  $ 0.1 \pm 0.2 $  \\
Period ratio, $P_c/P_b$ &  $\mathcal{N}_{\mathcal{U}}(a=2.0,b=2.06,\mu=2.03,\sigma=0.013)$ &  $ 2.02947 \pm 0.00011 $  \\
$ e_{c} $ &  $|\mathcal{N}(0,0.096)|$ &  $ 0.047^{+0.028}_{-0.024} $  \\
$ \omega_{c} $ &  $\mathcal{U}(a=-3.14,b=3.14)$ &  $ -0.5^{+2.1}_{-1.2} $  \\
Inclination, $ i_{c} $ [$^{\circ}$] &  $\mathcal{N}(\mu=1.5673,\sigma=0.0061)$ &  $ 1.5678^{+0.006}_{-0.0053} $  \\
Longitude of Ascending Node, $ \Omega_{c} $ [$^{\circ}$] &  $\mathcal{U}(a=-3.14,b=3.14)$ &  $ -1.2^{+2.7}_{-1.4} $  \\
Transit Epoch, $ t_{0, c} $ [BJD-2457000] &  $\mathcal{N}(\mu=1748.689,\sigma=0.05)$ &  $ 1748.697 \pm 0.013 $  \\
log mass ratio, $ \log{M_{p,d} / M_{p, b}} $ &  $\mathcal{N}_{\mathcal{U}}(a=-2.73,b=3.27,\mu=0.27,\sigma=0.62)$ &  $ 0.14^{+0.6}_{-0.5} $  \\
Period ratio, $P_d/P_b$ &  $\mathcal{N}_{\mathcal{U}}(a=3.3621,b=3.4221,\mu=3.3921,\sigma=0.0044)$ &  $ 3.39209 \pm 9e-05 $  \\
$ e_{d} $ &  $|\mathcal{N}(0,0.096)|$ &  $ 0.075 \pm 0.052 $  \\
$ \omega_{d} $ &  $\mathcal{U}(a=-3.14,b=3.14)$ &  $ -0.6^{+2.9}_{-1.2} $  \\
Inclination, $ i_{d} $ [$^{\circ}$] &  $\mathcal{N}(\mu=1.55166,\sigma=0.00099)$ &  $ 1.552 \pm 0.0011 $  \\
Longitude of Ascending Node, $ \Omega_{d} $ [$^{\circ}$] &  $\mathcal{U}(a=-3.14,b=3.14)$ &  $ -1.1^{+2.6}_{-1.7} $  \\
Transit Epoch, $ t_{0, d} $ [BJD-2457000] &  $\mathcal{N}(\mu=1762.667,\sigma=0.05)$ &  $ 1762.658 \pm 0.011 $  \\

\hline                                   %inserts single line
\end{tabular}
\end{table*}

\end{appendix}
\end{document}